\documentclass[a4paper,11pt]{article}
\usepackage{jcappub} % for details on the use of the package, please see the JINST-author-manual
\usepackage{lineno}
\arxivnumber{2401.01581} % Only if you have one
\title{\boldmath  Testing an exact diffraction formula with gravitational wave source lensed by a supermassive black hole in binary systems}

\author[a]{Xiao Guo}
\author[a,b,1]{and Zhoujian Cao\note{Corresponding author.}}
\emailAdd{zjcao@amt.ac.cn}
\affiliation[a]{School of Fundamental Physics and Mathematical Sciences, Hangzhou Institute for Advanced Study, University of Chinese Academy of Sciences, Hangzhou 310024, China}
\affiliation[b]{Institute of Applied Mathematics, Academy of Mathematics and Systems Science, Chinese Academy of Sciences, Beijing 100190, China}
\abstract{
When it comes to long-wavelength gravitational waves (GWs), diffraction effect becomes significant when these waves are lensed by celestial bodies. Typically, the traditional diffraction integral formula neglects large-angle diffraction, which is often adequate for most of cases. Nonetheless, there are specific scenarios, such as when a GW source is lensed by a supermassive black hole in a binary system, where the lens and source are in close proximity, where large-angle diffraction can play a crucial role. In our prior research, we have introduced an exact, general diffraction integral formula that accounts for large-angle diffraction as well. This paper explores the disparities between this exact diffraction formula and the traditional, approximate one under various special conditions. Our findings indicate that, under specific parameters — such as a lens-source distance of $D_{\rm LS}=0.1$\,AU and a lens mass of $M_{\rm L}=4\times10^{6}M_\odot$ — the amplification factor for the exact diffraction formula is notably smaller than that of the approximate formula, differing by a factor of approximately $r_F\simeq0.806$. This difference is substantial enough to be detectable. Furthermore, our study reveals that the proportionality factor $r_F$ gradually increases from 0.5 to 1 as $D_{\rm LS}$ increases, and decreases as $M_{\rm L}$ increases. Significant differences between the exact and approximate formulas are observable when $D_{\rm LS}\lesssim0.2$\,AU  (assuming $M_{\rm L}=4\times10^{6}M_\odot$) or when  $M_{\rm L}\gtrsim2\times10^6M_\odot$ (assuming $D_{\rm LS}=0.1$\,AU). These findings suggest that there is potential to validate our general diffraction formula through future GW detections.
}

\begin{document}
\maketitle
\flushbottom

\section{Introduction}
\label{sec:intro}

Gravitational lensing (GL)\cite{1992grle.book.....S,2006glsw.conf.....M} and gravitational waves (GW)\cite{1973grav.book.....M,maggiore2008gravitational} constitute two crucial predictions of General Relativity (GR). Both phenomena have been observed \cite{GW150914, 1992grle.book.....S,2006glsw.conf.....M}, thereby validating GR's predictions\cite{GWTC1_tgr,O3a_tgr}. Analogous to the GL of electromagnetic (EM) waves, GR also anticipates the occurrence of GL effect of GWs. However, up to now, no strongly lensed GW event has been definitively confirmed \cite{2021ApJ...923...14A}. Current observational efforts have identified several potential lensed GW event candidates\cite{2021ApJ...923...14A,2020arXiv200712709D}, but their confirmation remains challenging. Nevertheless, with the increasing detection of GW events, the discovery of GL events of GW events is imminent.

The detection of GL of GWs holds profound implications. It offers a unique opportunity to measure the propagation speed of GWs\citep{2017PhRvL.118i1101C,2017PhRvL.118i1102F}, constrain cosmological models\citep{2017NatCo...8.1148L}, and shed light on the origins of stellar binary black holes (BHs)\cite{2022ApJ...940...17C}. Furthermore, it provides a powerful tool to explore the nature of dark matter\cite{2018PhRvD..98j4029D,2020ApJ...901...58O,2020MNRAS.495.2002L,2021PhRvD.104f3001C,2022PhRvD.106b3018G,2022arXiv221211960T,2023JCAP...07..007F,2023arXiv230706990C,2023MNRAS.518..149Z} and even the internal structure of the sun\cite{2023JCAP...07..042J,2023ApJ...957...52T}. The potential applications of detecting GL of GWs are vast and promising, paving the way for exciting new discoveries in the field of gravitational physics.

Different from lensed EM wave, lensed GW poses unique challenges due to the potentially long wavelengths of detectable GWs, whose wavelength can approach or even exceed the scale of celestial objects \cite{Takahashi_Thesis}. This characteristic increases the likelihood of diffraction effects occurring in lensed GWs. Assuming a point mass lens model, diffraction effects become significant, when the lens mass is less than $10^5M_\odot\left(\frac{f}{\rm Hz} \right)^{-1}$ for GWs.

The traditional diffraction integral formula for GW lensing has been extensively studied and applied in various contexts\cite{1992grle.book.....S,1999PThPS.133..137N,Takahashi_Thesis}. However, this formula has a limitation: it neglects large-angle diffraction, which can lead to inaccuracies in specific scenarios. To address this issue, Guo and Lu \cite{2020PhRvD.102l4076G} introduced a more general diffraction integral formula. Although this exact formula typically yields results consistent with the traditional, approximate formula in most cases, it offers improved accuracy in those special instances where large-angle diffraction is significant.
In this paper, we aim to explore how to empirically test this exact formula in specific scenarios. We will compare the differences between the approximate and exact diffraction integral formulas, discussing whether these differences are detectable. Our investigation aims to provide a more comprehensive understanding of GW lensing and diffraction effects, paving the way for more accurate future studies in this field.

This paper is organized as follows. In Section~\ref{sec:the}, we review the theoretical framework of the general diffraction integral formula, and discuss the accuracy of the approximate diffraction integral. Then we investigate observable differences in Section~\ref{sec:diff}. Finally, some discussions on our results are summarized in Section~\ref{sec:dis} and we draw conclusions in Section~\ref{sec:con}. 
Throughout the paper, we adopt the geometrical unit system $G=c=1$.

%What is more, geometrical optics is the high frequency limit of wave optics. To study wave optics is also helpful for us to understand the nature of GL. 

\section{Theoretical framework}
\label{sec:the}

\begin{figure}
\centering
\includegraphics[width=0.5\textwidth]{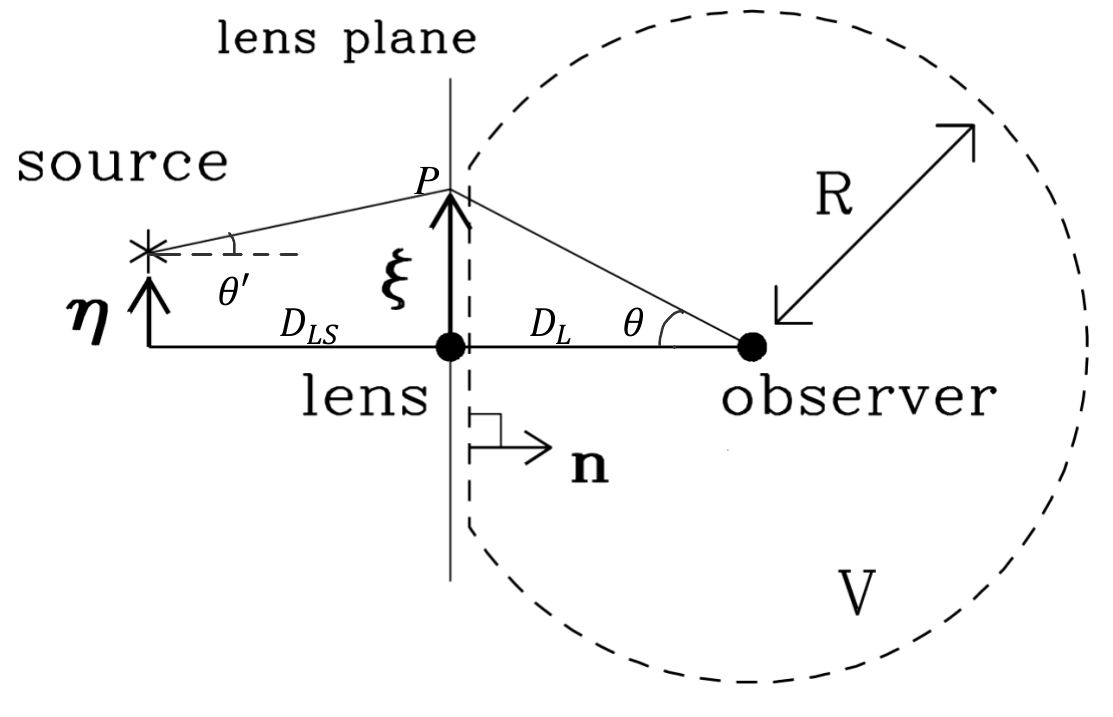}
\caption{
The schematic diagram to demonstrate the geometrical configuration of the observer-lens-source system (the same as \cite{2020PhRvD.102l4076G}). Here $D_{\rm L}$ ($D_{\rm S}$) represents the distance from the observer to the lens (source), and $D_{\rm LS}$ represents the distance between the lens and source; the distance between point $P$ and observer is $r$;  $\boldsymbol{n}$ is the normal vector of the lens plane; $\boldsymbol{\eta}$ ($\boldsymbol{\xi}$) is the position vector of GW source on the source plane (GW on the lens plane). 
}
\label{fig:illustr}
\end{figure}

For a GL system, its geometrical configuration is shown in Figure~\ref{fig:illustr}. Similar to \cite{2020PhRvD.102l4076G}, we adopt $D_{\rm S}$ ($D_{\rm L}$) to represent the distance between source (lens) and observer, and $D_{\rm LS}$ the distance between source and lens. The GW or EM wave emitted from a source at a position  $\boldsymbol{\eta}$ in the source plane reaches point $P$ or $\boldsymbol{\xi}$ on the lens plane and is finally received by the observer on Earth.  $\theta$ ($\theta'$) represents the angle between the normal vector of lens plane and GW propagation direction at the position of observer (source). 
%$r$ is the distance between point $P$ and observer.

It can be proven that the observed lensed waveform in the frequency domain $\tilde{\phi}^{L}$ propagating in a curved spacetime can be expressed as a surface integral at lens plane 
we usually use amplification factor to describe the lensing effect of this GL. The amplification factor is defined as the ratio of lensed waveform and unlensed waveform in the frequency domain: 
\begin{equation}
F(w,\boldsymbol{y})=\frac{\tilde{\phi}^{L}(w,\boldsymbol{y})}{\tilde{\phi}(w,\boldsymbol{y})},
\end{equation}
where $\omega$ represents the circular frequency of GW, 
$w=\frac{D_{\rm S}}{D_{\rm LS} D_{\rm L}} \xi_{0}^{2}\left(1+z_{\rm L}\right)\omega$ is a dimensionless frequency, $\boldsymbol{y}=\frac{D_{\rm L}}{\xi_{0} D_{S}} \boldsymbol{\eta}$ is dimensionless angular position of source, $\tilde{\phi}^{L}(w,\boldsymbol{y})$ and $\tilde{\phi}(w,\boldsymbol{y})$ are lensed and unlensed waveform of GW in the frequency domain, respectively and $\xi_0$ is a normalized constant of length, which is usually the Einstein radius of the lens.

Under the eikonal approximation, GW tensor or EM wave vector can be described as a scalar wave times constant basis. In usual cases of small angle diffraction, $\theta$, $\theta'\simeq0$, thus obliquity factor $\cos\theta$, $\cos\theta'\simeq1$. However, under general cases, e.g., large angle diffraction,  $\cos\theta$, $\cos\theta'\neq1$, hence the obliquity factor can not be ignored.
If we retain the obliquity factor in Kirchhoff diffraction integral formulas, it is easy to obtain (see details in \cite{2020PhRvD.102l4076G})
\begin{equation}
F(w, \boldsymbol{y})=\frac{w}{4 \pi i} \iint_S d^{2} x e^{i w T(\boldsymbol{x}, \boldsymbol{y})} (\cos\theta+\cos\theta') \cos\theta,
\label{eq:F_wycos}
\end{equation}
here $\cos\theta=\frac{\frac{D_{\rm L}}{\xi_0}}{\sqrt{\left(\frac{D_{\rm L}}{\xi_0}\right)^2+x^2}}$, $\cos\theta'=\frac{\frac{D_{\rm LS}}{\xi_0}}{\sqrt{\left(\frac{D_{\rm LS}}{\xi_0}\right)^2+\left(\boldsymbol{x}-\frac{D_{\rm S}}{D_{\rm L}}\boldsymbol{y}\right)^2}}$ and dimensionless time delay
\begin{equation}
T(\boldsymbol{x},\boldsymbol{y})=\frac{1}{2}|\boldsymbol{x}-\boldsymbol{y}|^{2}-\psi(\boldsymbol{x})+\phi_{m}(\boldsymbol{y}),
\label{eq:Time}
\end{equation}
with $\psi(\boldsymbol{x})$ representing the lens potential, $\phi_m(\boldsymbol{y})$ representing the arrival time in the unlensed case, $\boldsymbol{x}=\frac{\boldsymbol{\xi}}{\xi_{0}}$, $T(\boldsymbol{x}, \boldsymbol{y})=\frac{D_{\rm L} D_{\rm LS}}{D_{\rm S}\xi_{0}^{2}}  t_{\rm d}(\boldsymbol{\xi}, \boldsymbol{\eta})$, where $t_{\rm d}$ is the realistic lensing time delay of the GL system.
%

%scalar diffraction theory
What is more, this diffraction integral is not only applicable for GW or EM wave as an approximate formula, but also a scalar wave as an accurate formula.

\subsection{Axially symmetric case}
We adopt a polar coordinate system $(x,\theta_x)$ to describe this integral, where $x$ is the radius of coordinate point,
$\theta_x$ is the angle between $\boldsymbol{x}$ and $\boldsymbol{y}$, the positive direction of $\boldsymbol{y}$ is the polar axis. Thus Equation~\eqref{eq:F_wycos} can be rewritten as
\begin{equation}
F(w, \boldsymbol{y})=\frac{w}{4 \pi i} \int_0^{\infty} xd x\int_0^{2\pi} d\theta_x \exp\left[i w \left(\frac{x^2}{2}+\frac{y^2}{2}-xy\cos\theta_x-\psi(\boldsymbol{x})+\phi_{\rm m}(\boldsymbol{y})\right)\right] (\cos\theta+\cos\theta') \cos\theta,
\label{eq:F_wycospolar}
\end{equation}
where % \vec{y}  is constant (in dependent of x, \theta_x)  \cos\theta is independent of \theta_x 
$\cos\theta=\frac{\frac{D_{\rm L}}{\xi_0}}{\sqrt{\left(\frac{D_{\rm L}}{\xi_0}\right)^2+x^2}}$,
$
\cos\theta'=\frac{\frac{D_{\rm LS}}{\xi_0}}{\sqrt{\left(\frac{D_{\rm LS}}{\xi_0}\right)^2+x^2+\frac{D^2_{\rm S}}{D^2_{\rm L}}y^2-2\frac{D_{\rm S}}{D_{\rm L}}xy\cos\theta_x}}.
$
%under axial symmetric case.

Under axisymmetric case, we can show the explicit expression of Equation~\eqref{eq:F_wycos} in a simpler form. 
If the mass distribution of lens is axially symmetric, $\psi(\boldsymbol{x})=\psi(x)$ is independent of $\theta_x$, thus this equation is reduced to
\begin{equation}
\begin{aligned}
F(w, \boldsymbol{y})
=&\frac{w}{4 \pi i} e^{iw(y^2/2+\phi_m(\boldsymbol{y}))}\int_0^{\infty} xdx e^{iw(x^2/2-\psi(x))} \cos\theta\int_0^{2\pi} d\theta_x e^{-i w xy\cos\theta_x} (\cos\theta+\cos\theta').\\
%=&\frac{w}{i} e^{iw(y^2/2+\phi_m(\boldsymbol{y}))}\int_0^{\infty} xdx e^{iw(x^2/2-\psi(x))}J_0(wxy),
%
\end{aligned}
%
%\label{eq:axial}
%
\end{equation}

\subsection{Approximate form under small angle approximation}

If we adopt the small-angle approximation, i.e., $\cos\theta\simeq1$, $\cos\theta'\simeq1$, the distance between point $P$ and observer $r = D_{\rm L}/\cos \theta \simeq D_{\rm L}$, which holds true  for most astrophysical lensing systems, the amplification factor can be reduced to 
\begin{equation}
F(w, \boldsymbol{y})=\frac{w}{2 \pi i} \iint d^{2} x \exp [i w T(\boldsymbol{x}, \boldsymbol{y})].
\label{eq:F_wy}
\end{equation}
This formula is widely used in the calculation of amplification factor in the diffraction regime (e.g., see \citep{2003ApJ...595.1039T,2014PhRvD..90f2003C, 2018PhRvD..98j4029D, 2019ApJ...875..139L,2023MNRAS.524.2954M}). Under the axisymmetric case, we have
\begin{equation}
%
%\begin{aligned}
%
F(w, \boldsymbol{y})
=\frac{w}{i} e^{iw(y^2/2+\phi_m(\boldsymbol{y}))}\int_0^{\infty} xdx e^{iw(x^2/2-\psi(x))}J_0(wxy),
%
%\end{aligned}
%
\label{eq:axial}
\end{equation}
where $J_0(z)$ is zero-order Bessel function and it can be expressed as 
$$
J_0(z)=\frac{1}{\pi}\int_0^{\pi}e^{iz\cos\theta}d\theta. 
$$

\subsection{Accuracy of the small angle approximation}
\label{sec:acc}

Since $\frac{D_{\rm L}}{\xi_0}$ and $\frac{D_{\rm LS}}{\xi_0}$ are usually large numbers under most of cases, while $x$, $y$ are relatively small, $\frac{D_{\rm L}}{\xi_0}$, $\frac{D_{\rm LS}}{\xi_0}\gg x, y$, $\cos\theta$ and $\cos\theta'$ can be expanded in the form of series and we have
\begin{equation}
\frac{(\cos\theta+\cos\theta')\cos\theta}{2}\approx1-\frac{3}{4}\left(\frac{x}{\frac{D_{\rm L}}{\xi_0}}\right)^2-\frac{1}{4}\left(\frac{\boldsymbol{x}-\frac{D_{\rm S}}{D_{\rm L}}\boldsymbol{y}}{\frac{D_{\rm LS}}{\xi_0}}\right)^2+...
\end{equation}
If we regard Equation~\eqref{eq:F_wycos} as the corrected results, this expansion can also give an estimate to the relative error of the traditional form Equation~\eqref{eq:F_wy}, which is roughly
\begin{equation}
\epsilon=\mathcal{O}\left(\left(\frac{x}{\frac{D_{\rm L}}{\xi_0}}\right)^2\right)+\mathcal{O}\left(\left(\frac{|\boldsymbol{x}-\frac{D_{\rm S}}{D_{\rm L}}\boldsymbol{y}|}{\frac{D_{\rm LS}}{\xi_0}}\right)^2\right).
\label{eq:epsilon}
\end{equation}
The normalization length $\xi_0$ is usually taken as Einstein radius $\xi_0=r_{\rm E}=\sqrt{2R_{\rm S}\frac{D_{\rm LS}D_{\rm L}}{D_{\rm S}}}$, where $R_{\rm S}$ the Schwarzschild radius of the lens object. 
As discussed in \cite{2020PhRvD.102l4076G}, for the usual faraway lens, it is difficult to detect the tiny difference between exact and approximate diffraction formula. Thus we consider some special cases: the sun as the lens or the lens-source  binary system.

\subsubsection{Solar lens}
\label{sec:sun}
The sun is nearest massive lens to us. Similar to \cite{2023JCAP...07..042J,2023ApJ...957...52T}, we choose the sun as the lens. In this case, we adopt some typical parameters $D_{\rm L}=1$\,AU=$1.5\times10^8$\,km, $D_{\rm LS}\sim D_{\rm S}\gg D_{\rm L}$, source (e.g. pulsar \cite{2023ApJ...957...52T}) distance $D_{\rm S}\sim 1$\,kpc=$3\times10^{16}$\,km (see pulsar catalog\footnote{http://www.atnf.csiro.au/people/pulsar/psrcat} \citep{2005AJ....129.1993M}), Schwarzschild radius $R_{\rm S}=3$\,km for sun, and $\xi_0\simeq\sqrt{2R_{\rm S}D_{\rm L}}$.
We can estimate
$$
\frac{D_{\rm L}}{\xi_0}=\sqrt{\frac{D_{\rm L}}{2R_{\rm S}}}\sim 5000\gg 1,
$$
thus the first term in Equation~\eqref{eq:epsilon} is a small quantity.
And
$$
\frac{|\boldsymbol{x}-\frac{D_{\rm S}}{D_{\rm L}}\boldsymbol{y}|}{\frac{D_{\rm LS}}{\xi_0}}\simeq\frac{|\boldsymbol{x}-\frac{D_{\rm S}}{D_{\rm L}}\boldsymbol{y}|}{\sqrt{\frac{D_{\rm S}^2}{2R_{\rm S}D_{\rm L}}}}\simeq\frac{|\boldsymbol{x}-2\times10^{8}\boldsymbol{y}|}{10^{12}}\ll1
$$
for usual $\boldsymbol{x}, \boldsymbol{y}\sim O(1)$,
thus the second term in Equation~\eqref{eq:epsilon} is also a small quantity. Therefore, the traditional form Equation~\eqref{eq:F_wy} for diffraction integral is applicable and accurate even for solar lens.

\subsubsection{Lens-source binary system}
\label{sec:binary}
Next, we consider a scenario similar to that explored in \citep{2020PhRvD.101b4039M}: when the lens and source constitute a binary system, and the source is eclipsed by the lens. In this situation, three distances have such relation $D_{\rm LS}\ll D_{\rm L}\simeq D_{\rm S}$. It's important to clarify that the lens is not the entire binary system, but rather one component of it, with the source being the other component. This binary system could be composed of various combinations such as binary neutron stars (NS), NS-black hole (BH) binaries, compact object-star binaries, and so forth. The lens might be a BH or another compact object, while the gravitational wave source could originate from a rotating asymmetric NS or a binary system.
In this particular case, the binary system's distance from us is approximately $D_{\rm L}\simeq D_{\rm S}\sim1$\,kpc = $3\times10^{16}$\,km. The distance between the lens and source within the binary system is much smaller, $D_{\rm LS}\ll D_{\rm L}$ and $D_{\rm LS}\sim 1$\,AU=$1.5\times10^8$\,km. Assuming the lens has a mass comparable to the solar mass ($M_{\rm L}\simeq1M_\odot$), its Schwarzschild radius $R_{\rm S}=3$\,km and the normalization length $\xi_0\simeq\sqrt{2R_{\rm S}D_{\rm LS}}$.
%Then we consider the second case (similar to \citep{2020PhRvD.101b4039M}): if the lens and source form a binary system, source is eclipsed by the lens, thus $D_{\rm LS}\ll D_{\rm L}\simeq D_{\rm S}$. We need to stress that, the lens is not the binary system, but one component in a binary system, and the another one component in this system is source. This binary system could be binary neutron stars (NS), NS-BH binary system, compact object-star binary and so on. The lens could be a BH or other compact object and the GW source could be a rotating asymmetric NS or binary system.  In this case, binary distance to us is $D_{\rm L}\simeq D_{\rm S}\sim1$\,kpc = $3\times10^{16}$\,km, $D_{\rm LS}\ll D_{\rm L}$, distance between binary $D_{\rm LS}\sim 1$\,AU=$1.5\times10^8$\,km, Schwarzschild radius $R_{\rm S}=3$\,km for the lens (if its mass is around solar mass $M_{\rm L}\simeq1M_\odot$) and normalization length $\xi_0\simeq\sqrt{2R_{\rm S}D_{\rm LS}}$.
Similarly, we can estimate
$$
\frac{D_{\rm L}}{\xi_0}=\sqrt{\frac{D_{\rm L}^2}{2R_{\rm S}D_{\rm LS}}}\sim10^{12}\gg1,
$$
thus the first term in Equation~\eqref{eq:epsilon} is also a small quantity.
The second term
$$
\frac{|\boldsymbol{x}-\frac{D_{\rm S}}{D_{\rm L}}\boldsymbol{y}|}{\frac{D_{\rm LS}}{\xi_0}}=\frac{|\boldsymbol{x}-\frac{D_{\rm S}}{D_{\rm L}}\boldsymbol{y}|}{\sqrt{\frac{D_{\rm LS}}{2R_{\rm S}}}}\simeq\frac{|\boldsymbol{x}-\boldsymbol{y}|}{5000}
$$
is also a small quantity for $\boldsymbol{x}, \boldsymbol{y}\sim O(1)$.

However, if the lens becomes extremely massive, such as $M_{\rm L}\sim10^6M_\odot$, the second term can approach or even exceed 1, when $\boldsymbol{x}, \boldsymbol{y}\sim O(1)$. Consider, for instance, a supermassive BH acting as the lens \cite{2021SCPMA..6420462Z}, which could potentially be located in the center of our Galaxy or other nearby galaxies. We summarize the distance ratios for various scenarios, including supermassive BHs with different parameters, in Table~\ref{tab:D_ratio}.
This table presents estimation results not only for the supermassive BH in our Galactic center but also for nearby galactic centers, including M31, M32, M87, and others \cite{2022ApJ...939...55G}. It becomes evident from the table that only when the lens is a supermassive BH and the condition $D_{\rm LS}\ll D_{\rm L}\simeq D_{\rm S}$ is met, the ratio $\frac{D_{\rm S}}{D_{\rm L}}/\frac{D_{\rm LS}}{\xi_0}$ becomes $O(1)$ or even exceeds 1. In these specific cases, while the approximation $\cos\theta\simeq1$ remains valid, $\cos\theta'\simeq1$ is no longer accurate. Consequently, the traditional form of the diffraction integral given by Equation~\eqref{eq:F_wy} becomes inadequate and may lead to inaccuracies. To obtain more precise results, we must adopt the exact and more general formulas, namely Equation~\eqref{eq:F_wycos} or \eqref{eq:F_wycospolar}, for calculating the amplification factor in a binary system comprising a supermassive BH as the lens and a source.

%However, if the lens become very massive, e.g. $M_{\rm L}\sim10^6M_\odot$, the second term could become a value that is close to 1 even greater than 1 for $\boldsymbol{x}, \boldsymbol{y}\sim O(1)$.  For example, the lens is a supermassive BH\cite{2021SCPMA..6420462Z}, and this supermassive BH could be located in our Galactic center or other nearby galactic center. We summarize all these distance ratios for more cases (including supermassive BH) with different parameters in Table~\ref{tab:D_ratio}. In this table, we not only show the estimation results about supermassive BH in our Galactic center, but also show the estimation results for nearby Galactic center including M31, M32, M87 and so on\cite{2022ApJ...939...55G}.   From this table, we can see that only if when the lens is a supermassive BH and $D_{\rm LS}\ll D_{\rm L}\simeq D_{\rm S}$, ratio $\frac{D_{\rm S}}{D_{\rm L}}/\frac{D_{\rm LS}}{\xi_0}$ becomes $O(1)$ even greater than 1. In these cases, $\cos\theta\simeq1$ is still accurate, but $\cos\theta'\simeq1$ is no longer accurate. Hence the traditional form Equation~\eqref{eq:F_wy} for diffraction integral is not applicable and accurate. To obtain more accurate results, we need to adopt new, general formula Equation~\eqref{eq:F_wycos} or \eqref{eq:F_wycospolar} for diffraction integral to calculate amplification factor in the binary system consisting of a supermassive BH (as lens) and source. 

\begin{table}[htbp]
\centering
\caption{These distance ratios for different lens-source systems with different parameters in different cases. GC: Galactic center, M31C/M32C/M87C: the center of M31/M32/M87.}
\begin{tabular}{cc|cccc|cccc}
\hline
& Case & $M_{\rm L}/M_\odot$ & $D_{\rm L}$ & $D_{\rm LS}$ & $D_{\rm S}$  &$\frac{D_{\rm L}}{\xi_0}$ & $\frac{D_{\rm S}}{D_{\rm L}}$&$\frac{D_{\rm LS}}{\xi_0}$ &$\frac{D_{\rm S}}{D_{\rm L}}/\frac{D_{\rm LS}}{\xi_0}$\\
\hline
(1) & Sun & 1 & 1AU & 1kpc & 1kpc & $\gg 1$&$2\times10^8$ & $10^{12}$ &$\ll 1$\\
(2) & Binary & 1 & 1kpc & 1AU & 1kpc & $\gg 1$& 1 & $\gg1$& $\ll1$ \\
\hline
(3) & GC& $4\times 10^6$ & 8kpc& 1AU & 8kpc& $\gg 1$& 1 & 2.5 & 0.4 \\
(4)& GC& $4\times 10^6$ & 8kpc& 0.1AU & 8kpc& $\gg 1$& 1 & 0.79 &  1.27\\
(5)& GC& $4\times 10^6$ & 8kpc& 8kpc & 16kpc &  $\gg 1$& 2 &  $\gg 1$ & $\ll1$ \\
\hline
(6) &M31C & $1.4\times 10^8$ & 774kpc& 1AU & 774kpc& $\gg 1$& 1&0.42& 2.38\\
(7) &M32C & $2.4\times 10^6$ & 805kpc& 1AU & 805kpc& $\gg 1$& 1&3.23& 0.31\\
(8) &M87C & $6.5\times 10^9$ & 14.6Mpc& 1AU & 14.6Mpc& $\gg 1$& 1&0.062&16.1\\
\hline
\end{tabular}
\label{tab:D_ratio}
\end{table}
\section{Observable Differences}
\label{sec:diff}

Since the difference between exact and approximate diffraction integral formula is significant in the binary system consisting of a supermassive BH (as lens) and source, we  discuss the observable differences between them from several different aspects as below for the special case. 

\subsection{$F$ as the function of frequency $w$}
\label{sec:F_w}

\begin{figure}[htbp]
\centering
\includegraphics[width=.4\textwidth]{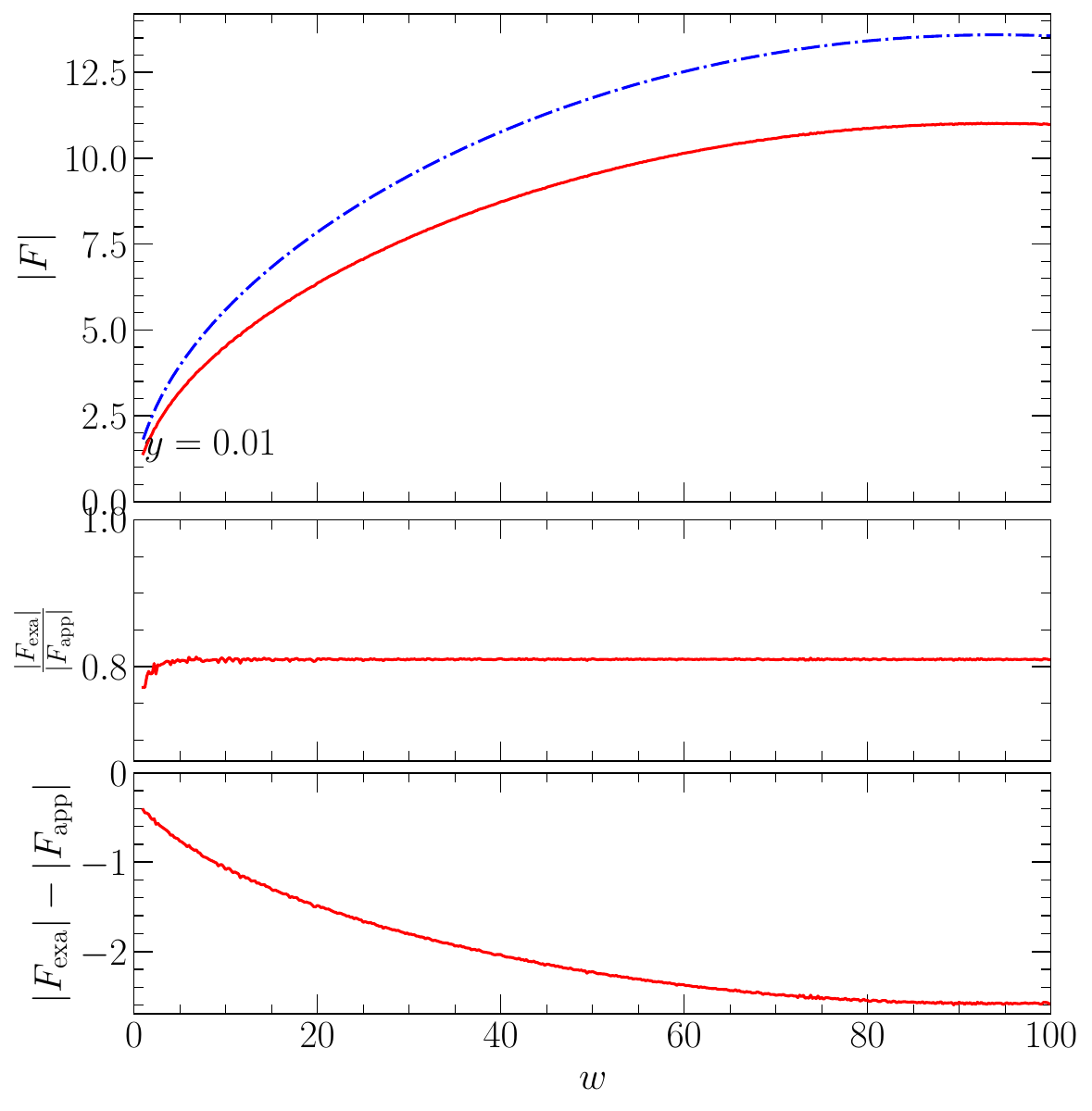}%GLnew/Plot.ipynb  &   F_Levin.nb
\includegraphics[width=.4\textwidth]{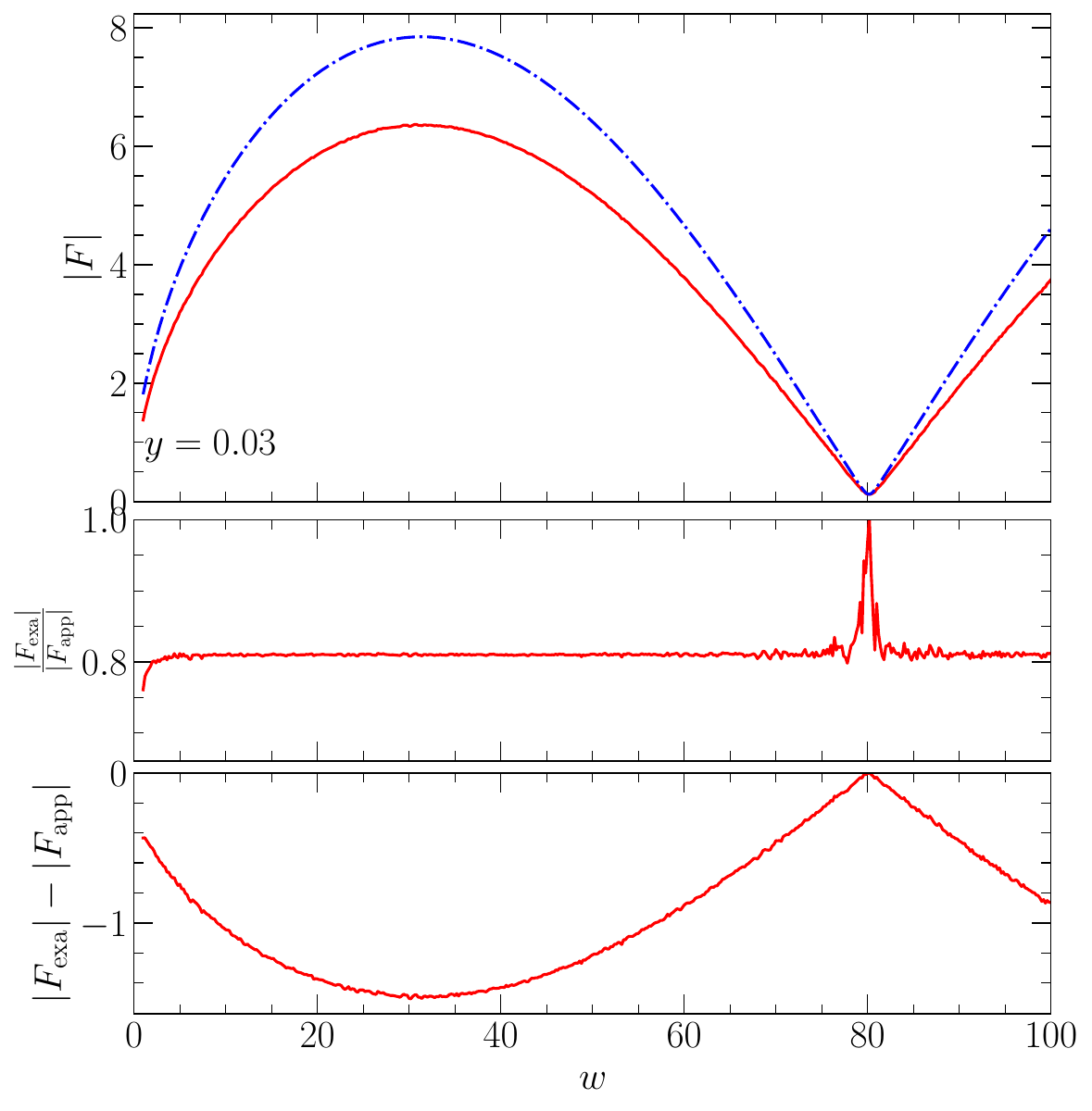}%GLnew/Plot.ipynb  &   F_Levin.nb
\\
\includegraphics[width=.4\textwidth]{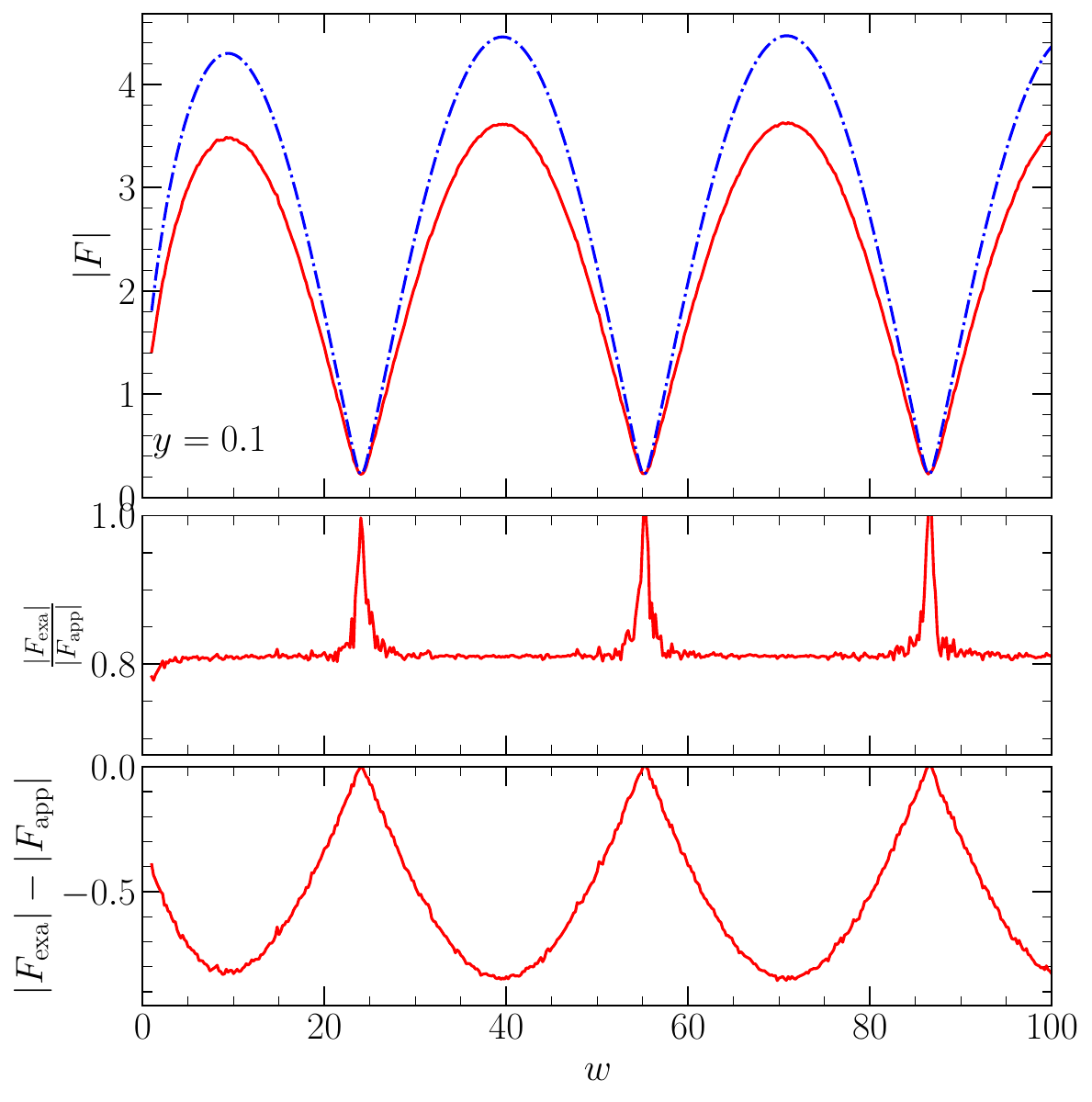}%GLnew/Plot.ipynb  &   F_Levin.nb
\includegraphics[width=.4\textwidth]{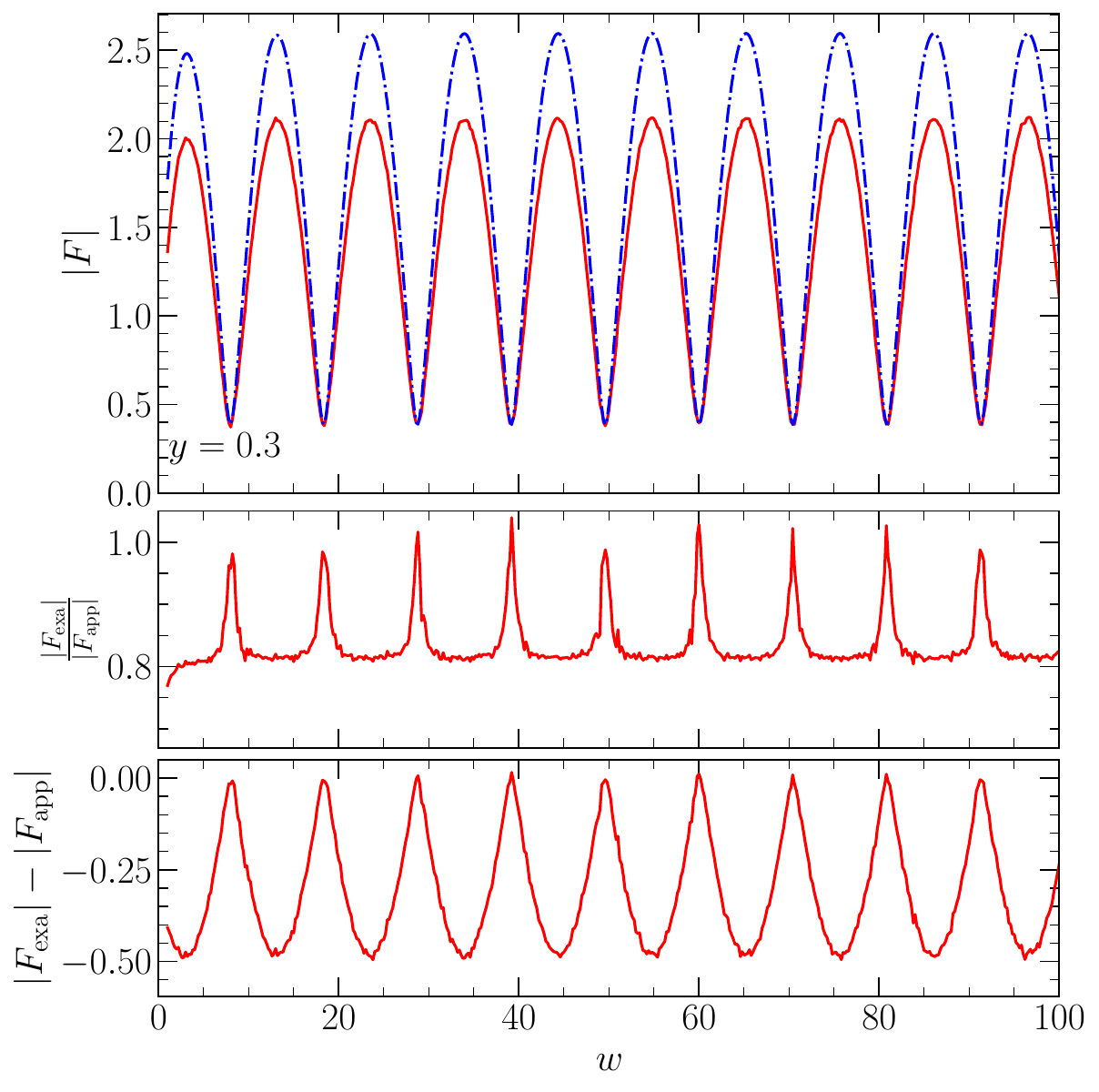}%GLnew/Plot.ipynb  &   F_Levin.nb
\caption{$|F|$ as the function of $w$ with different $y$ values, where the blue dot-dash (red solid) lines present results from approximate (exact) formula in top panel of each subfigure, while other panels reflects the difference between exact and approximate formula. 
The small-amplitude-oscillation of red curves could be due to numerical calculation error.}
\label{fig:F_w}
\end{figure}

For simplicity, we utilize the point mass lens model, which can describe black holes or other compact objects, as our lens model. For the point mass lens model, the lens potential is given by $\psi(x)=\ln x$, which is an axially symmetric lens. Therefore, we can employ Equation~\eqref{eq:F_wycospolar} to compute the exact diffraction integral.

Regarding numerical integration, certain traditional methods (e.g., \cite{1995ApJ...442...67U}) are unsuitable for the exact diffraction integral. Instead, we can use the Levin-type method \cite{Levin1982,2008mgm..conf..807M} or the asymptotic expansion method \cite{Takahashi_Thesis} for calculations (for more numerical techniques, see \cite{2020PhRvD.102l4076G}; for the Picard-Lefschetz method, see \cite{2019arXiv190904632F}).

If the GW source's frequency $f$ varies over time, it can sweep over a wide frequency range. Consequently, in Figure~\ref{fig:F_w}, we illustrate the amplification factor $F(w,y)$ as a function of the dimensionless frequency $w$ for different values of $y$. These calculations assume the same parameters as those listed in the fourth row (specifically, $D_{\rm LS}=0.1$\,AU) of Table~\ref{tab:D_ratio}. For clarity, Figure~\ref{fig:F_w} only displays the modules of $F$ for both exact and approximate formulas, as the phase difference between them is insignificant (see Appendix~\ref{sec:phase} for an example of the phase difference). Furthermore, since the absolute phase of lensing can be challenging to measure or calculate and is dependent on a specific basis, this paper primarily focuses on presenting the modules of amplification factors $|F|$.

In Figure~\ref{fig:F_w}, each subfigure corresponds to a different $y$ value, indicated at the bottom left corner of the upper panel in each respective subfigure. Across all subfigures, it's evident that $|F|$ exhibits oscillatory behavior with respect to $w$, particularly noticeable at higher $w$ values. The oscillation amplitude remains relatively constant. The blue dot-dashed line represents $|F_{\rm app}|$ computed using an approximate diffraction integral (Eq.~\ref{eq:F_wy}), while the red solid line signifies $|F_{\rm exa}|$ determined through an exact diffraction integral (Eq.~\ref{eq:F_wycos}). Notably, the red line consistently lies above the blue line. The ratios of $|F_{\rm exa}|$ and $|F_{\rm app}|$ are displayed in the middle panels of each subfigure. In the bottom panels, we show the difference $|F_{\rm exa}|-|F_{\rm app}|$. It's observable that $\frac{|F_{\rm exa}|}{|F_{\rm app}|}$ slightly increases with $w$ when $w\lesssim5$ and remains relatively constant thereafter, except at specific nodes. Since the ratio $\frac{|F_{\rm exa}|}{|F_{\rm app}|}$ is not constant, the amplification factor derived from the exact diffraction formula cannot be accurately fitted by merely adjusting parameters in the approximate formula. At these nodes, both $|F_{\rm exa}|$ and $|F_{\rm app}|$ attain local minima, while the ratio $\frac{|F_{\rm exa}|}{|F_{\rm app}|}$ reaches local maxima, and the absolute difference $\left||F_{\rm exa}|-|F_{\rm app}|\right|$ hits local minima.

\subsection{$F$ as the function of $y$}
\label{sec:F_y}

\begin{figure}[htbp]
\centering 
\includegraphics[width=.4\textwidth]{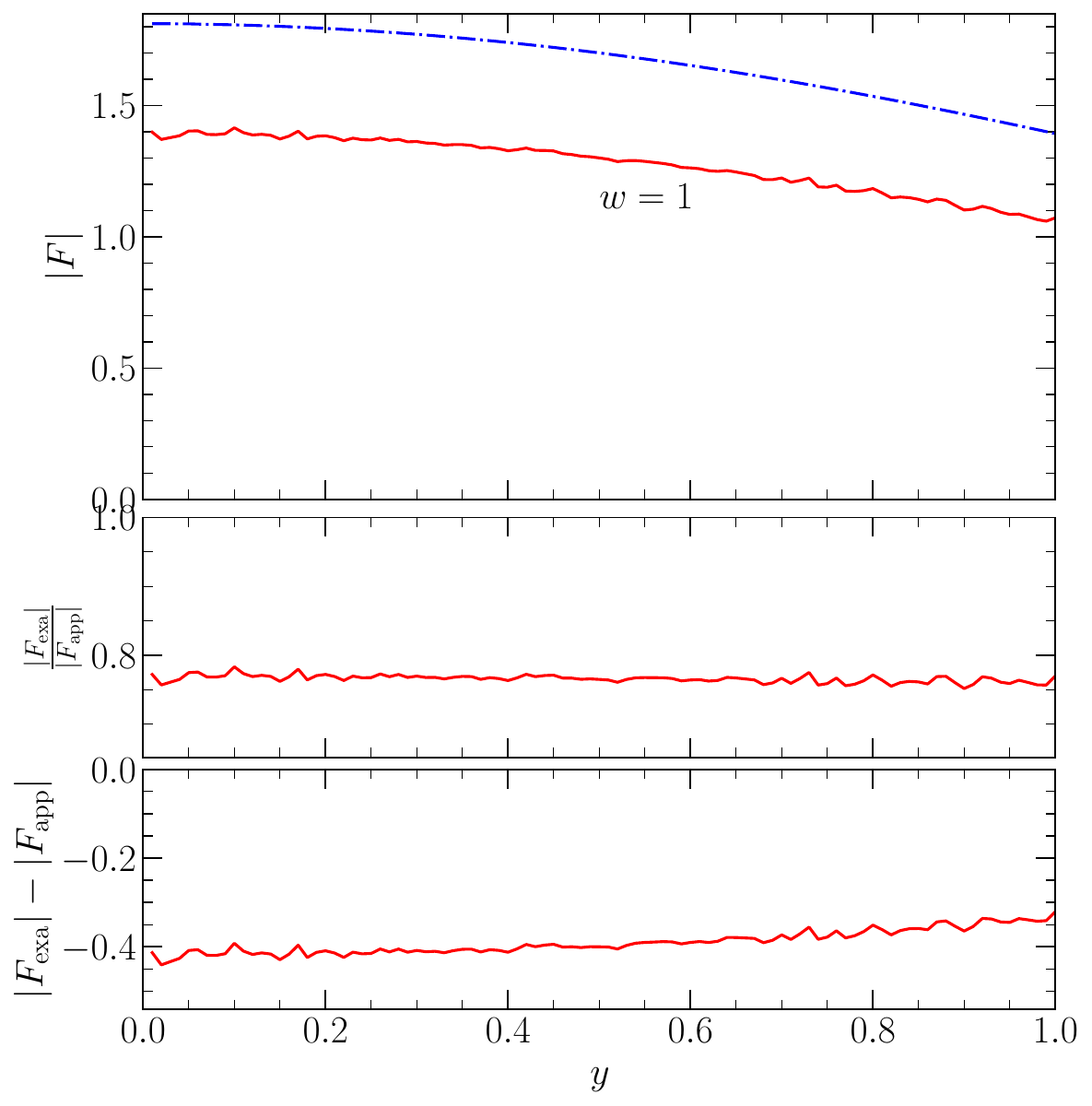}%GLnew/Plot.ipynb  &   F_Levin0.1lin.nb
\includegraphics[width=.4\textwidth]{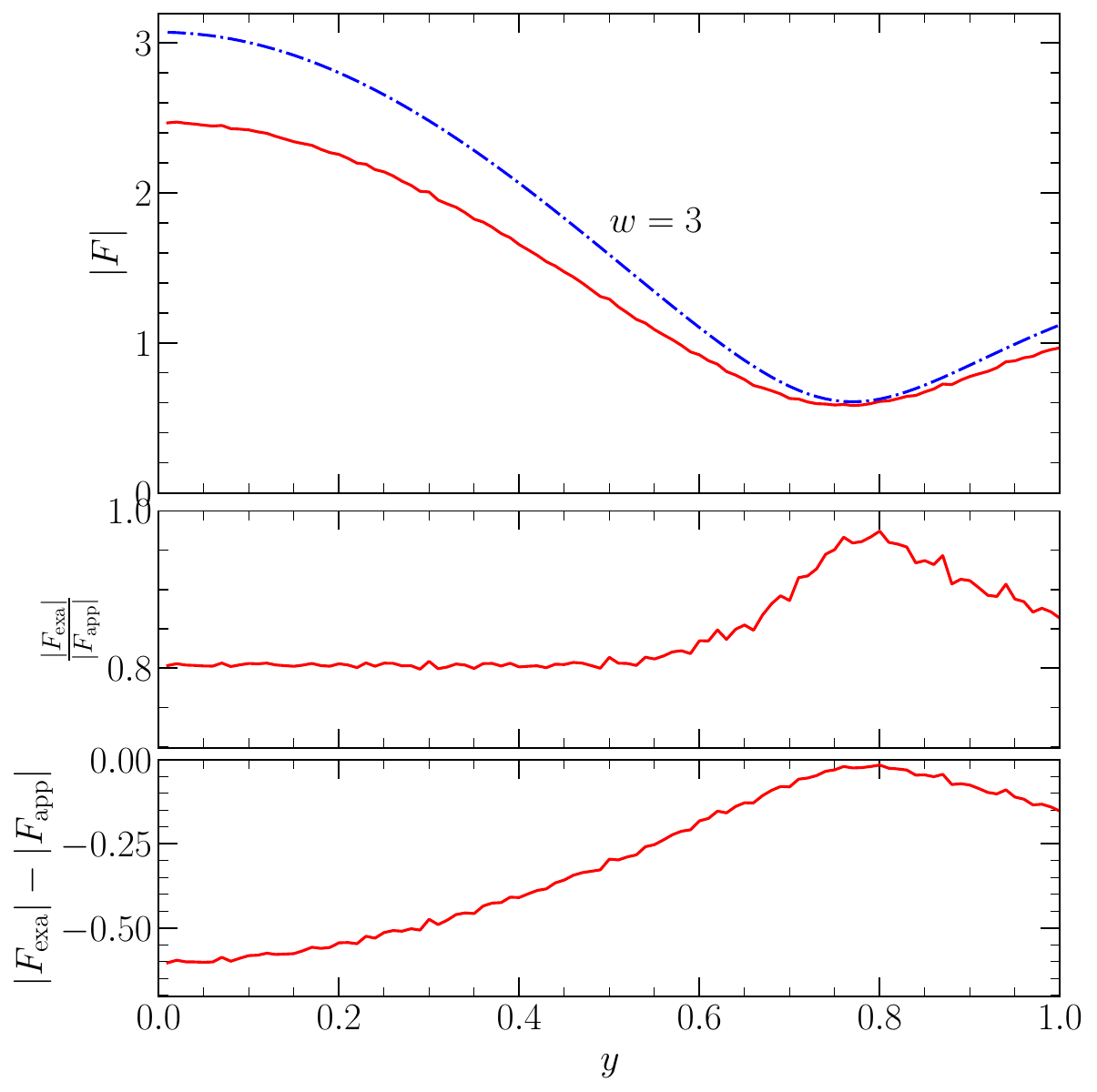}%GLnew/Plot.ipynb  &   F_Levin.nb
\\
\includegraphics[width=.4\textwidth]{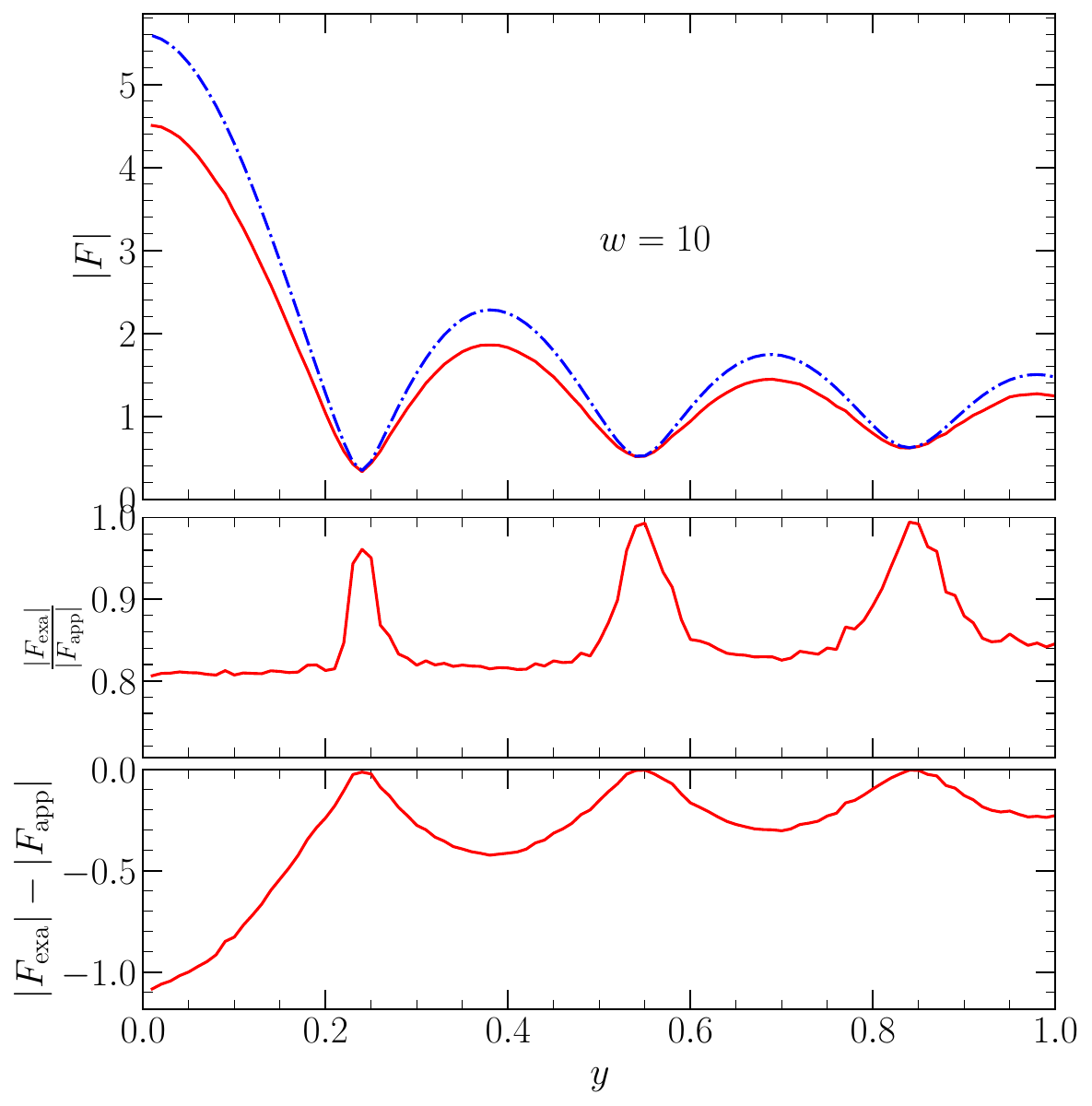}%GLnew/Plot.ipynb  &   F_Levin.nb
\includegraphics[width=.4\textwidth]{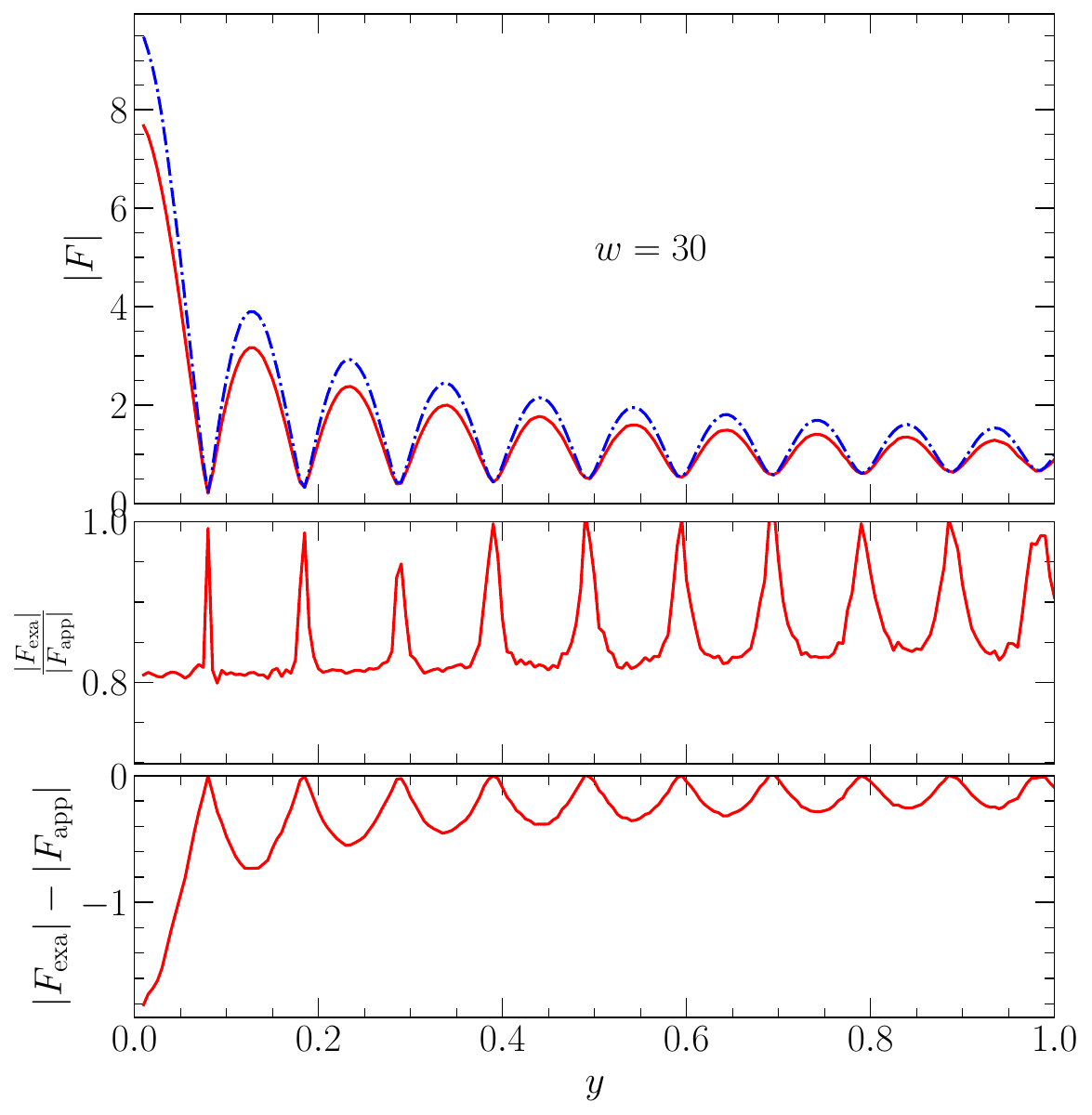}%GLnew/Plot.ipynb  &   F_Levin.nb
\caption{$|F|$ as the function of $y$ with different $w$ values, where the legends are similar to Figure~\ref{fig:F_w}.}
\label{fig:F_y}
\end{figure}

If the GW frequency remains constant, as is the case with a continuously emitting, uniformly rotating NS, the GW frequency, related to $w$, will be constant. However, if the GW source is in motion, the variable $y$ representing its position will be varying, thereby influencing the amplification factor $F$, which varies accordingly. To illustrate this relationship, we present the amplification factor $F(w,y)$ as a function of the dimensionless source position $y$ in Figure~\ref{fig:F_y}. This representation assumes the same parameters as those listed in the (4)th-row (specifically, $D_{\rm LS}=0.1$\,AU) of Table~\ref{tab:D_ratio}.

Within Figure~\ref{fig:F_y}, each subfigure corresponds to a distinct $w$ value, clearly indicated at the center of the upper panel in each respective subfigure. The legends in this figure are comparable to those in Figure~\ref{fig:F_w}. The blue dot-dashed line signifies $|F_{\rm app}|$ computed using an approximate diffraction integral, while the red solid line represents $|F_{\rm exa}|$ determined through an exact diffraction integral. Across all subfigures, it's evident that $|F|$ exhibits oscillatory behavior with respect to $y$, particularly noticeable at higher $y$ values. Nonetheless, the oscillation amplitude diminishes as $y$ increases, peaking when $y=0$. Notably, the red line consistently lies above the blue line. In the central panel of each subfigure, it's apparent that $\frac{|F_{\rm exa}|}{|F_{\rm app}|}$ remains relatively constant, except at specific nodes (attributable to $w\geq1$). At these nodes, both $|F_{\rm exa}|$ and $|F_{\rm app}|$ hit local minima, whereas the ratio $\frac{|F_{\rm exa}|}{|F_{\rm app}|}$ attains local maxima. Concurrently, the absolute difference $\left||F_{\rm exa}|-|F_{\rm app}|\right|$ reaches local minima.

%If the frequency of GW is constant, for example, the GW source is a uniformly rotating NS with emitting continuous GW. In this case, the frequency $w$ is constant, but if the GW source is moving, $y$ is varying, thus amplification factor $F$ is changing with $y$. Thus we show the amplification factor $F(w,y)$ as the function of dimensionless source position $y$ in Figure~\ref{fig:F_y} by assuming the same parameters as the (4)th-row (with $D_{\rm LS}=0.1$\,AU) in Table~\ref{tab:D_ratio}. 

%In Figure~\ref{fig:F_y}, each subfigure represents a different $w$, which is marked at the center of upper panel in each subfigure. All legends are similar to Figure~\ref{fig:F_w}. The blue dot-dash (red solid) line represents $|F_{\rm app}|$ ($|F_{\rm exa}|$) calculated by approximate (exact) diffraction integral. In each subfigure, we can see that the $|F|$ is also oscillating with $y$, especially when $y$ is high. But the oscillation amplitude is decreasing with increasing $y$ and it becomes maximum when $y=0$. The red line is also significantly higher than the blue line. In the middle panel of each subfigure, we can see that $\frac{|F_{\rm exa}|}{|F_{\rm app}|}$ is also basically a constant except some nodes (since $w\geq1$). At these nodes, $|F_{\rm exa}|$ and $|F_{\rm app}|$ reach local minimums, while the ratio $\frac{|F_{\rm exa}|}{|F_{\rm app}|}$ reaches local maximums, and the absolute value of difference $\left||F_{\rm exa}|-|F_{\rm app}|\right|$ reaches local minimums.

\subsection{Light curve examples}
\label{sec:F_t}

\begin{figure}[htbp]
\centering 
\includegraphics[width=.4\textwidth]{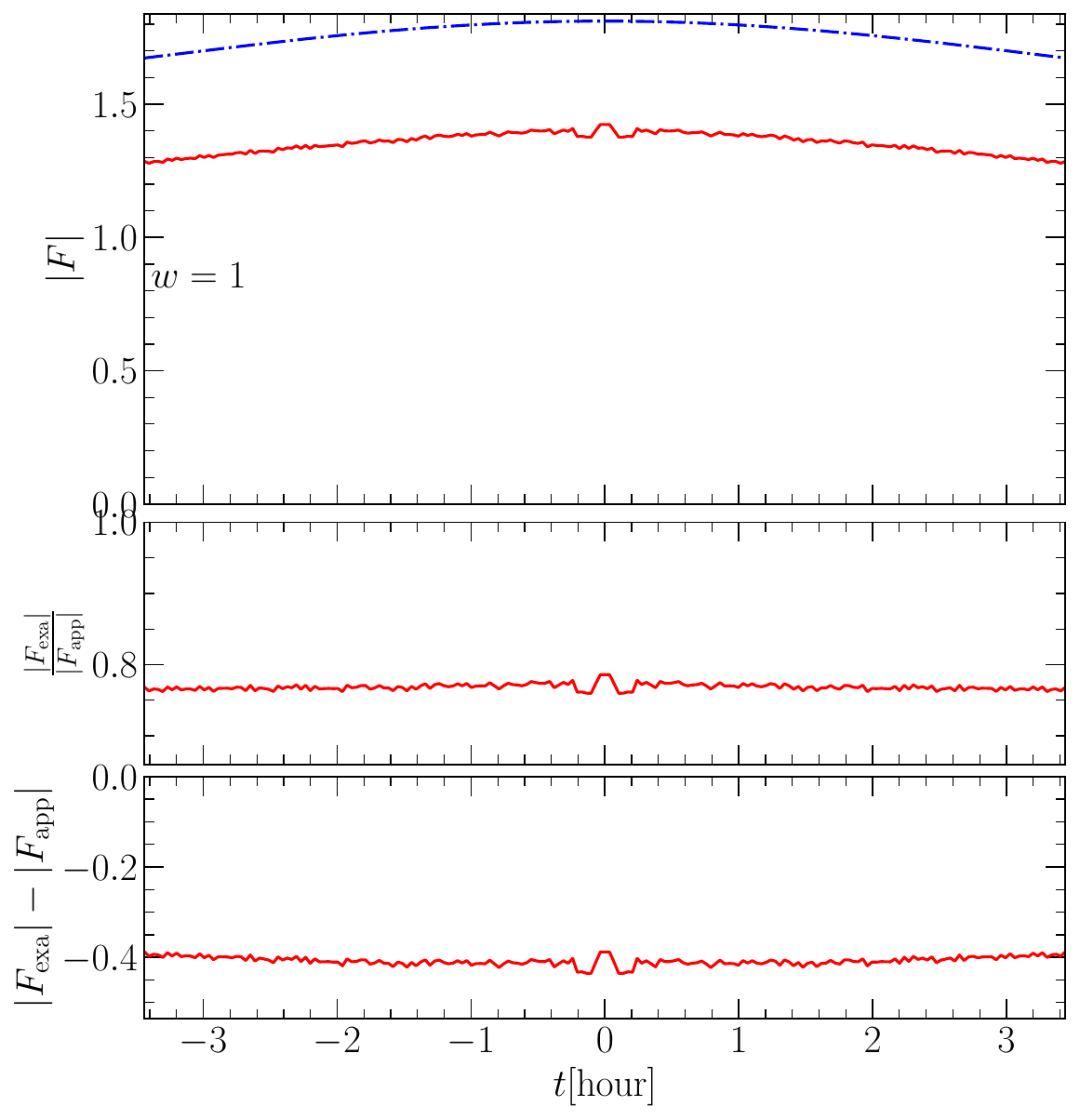}%GLnew/Plot.ipynb  &   F_Levin0.1lin.nb
\includegraphics[width=.4\textwidth]{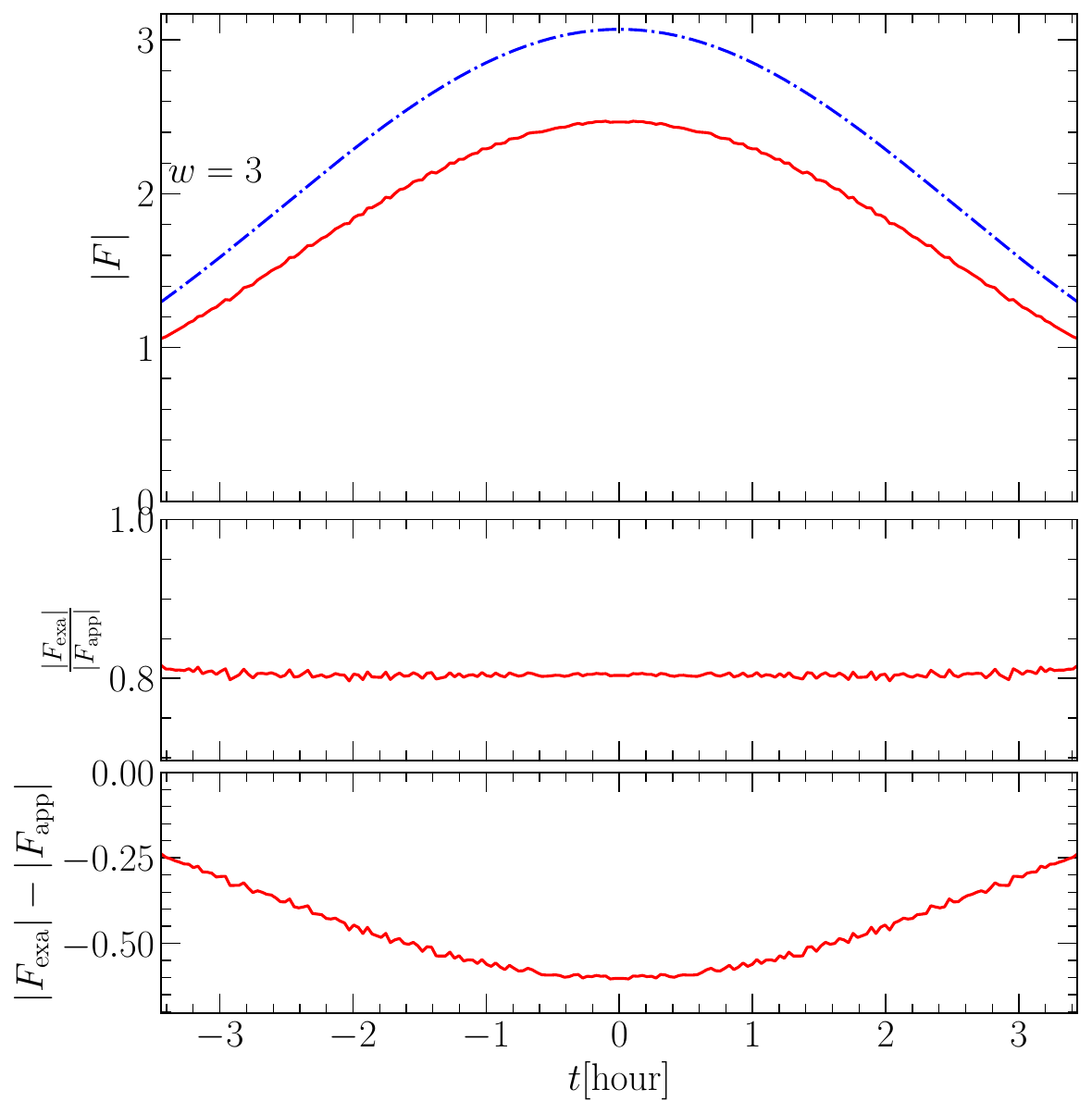} \\%GLnew/Plot.ipynb  &   F_Levin.nb
\includegraphics[width=.4\textwidth]{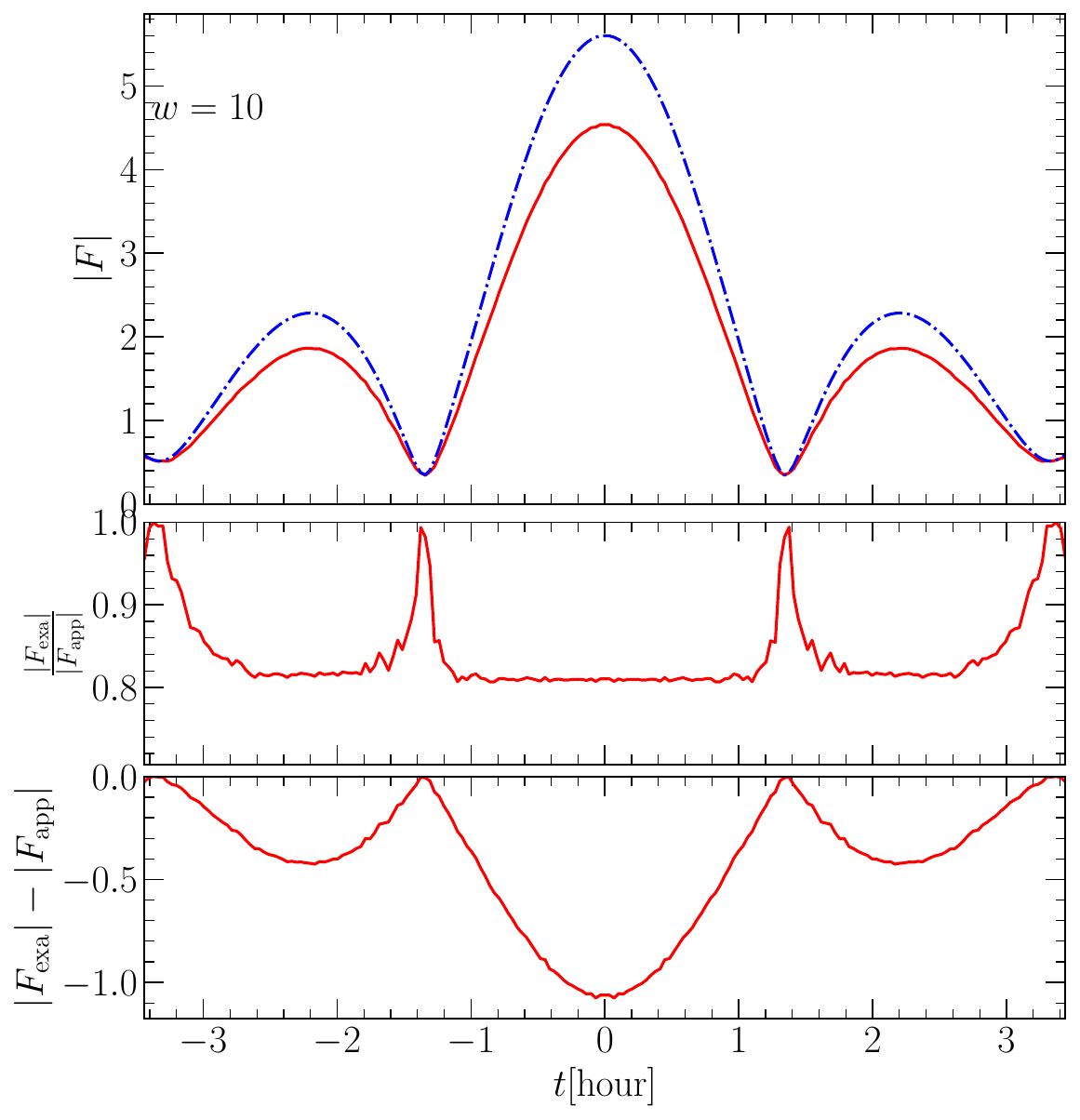}%GLnew/Plot.ipynb  &   F_Levin.nb
\includegraphics[width=.4\textwidth]{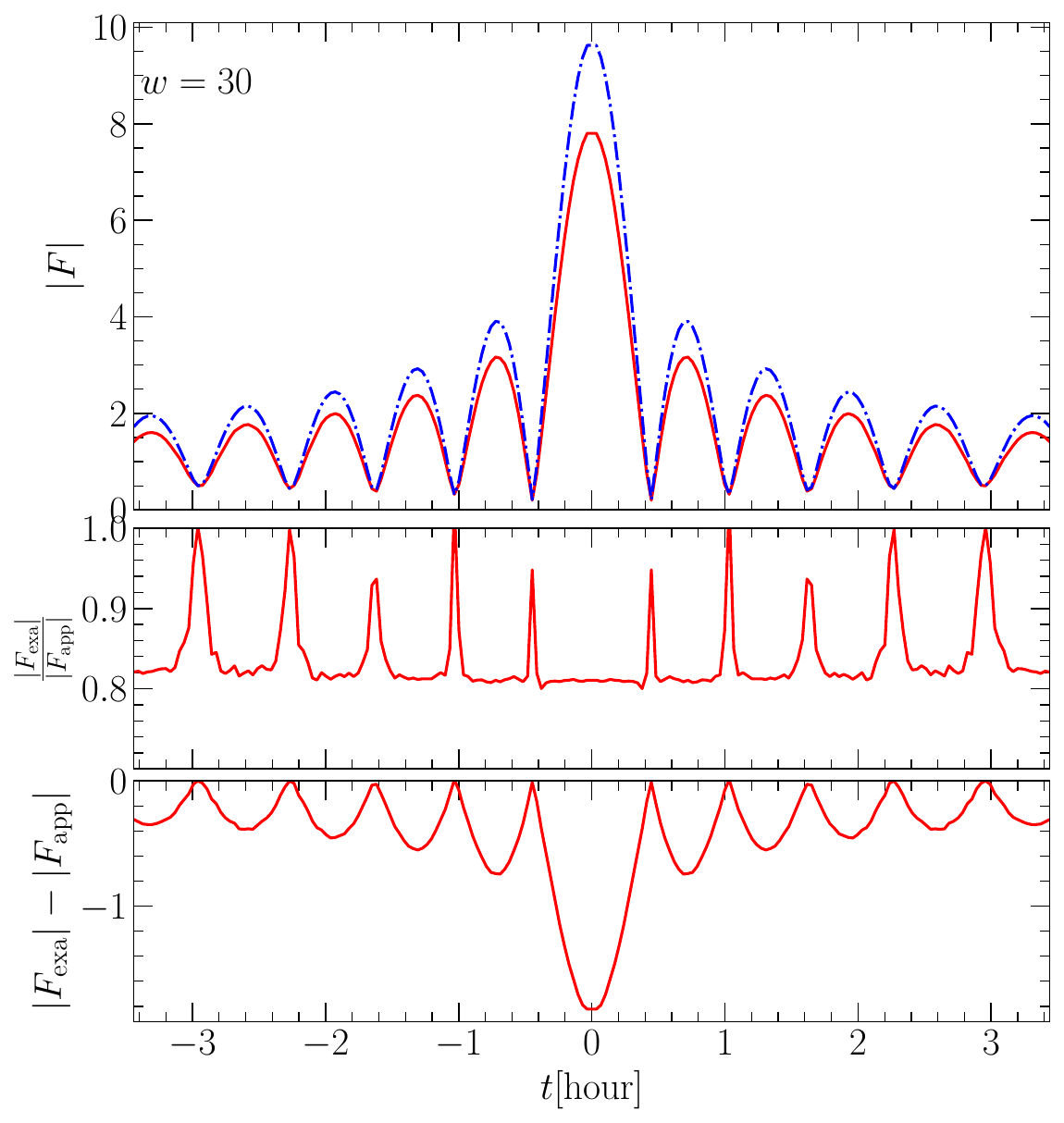}%GLnew/Plot.ipynb  &   F_Levin.nb
%\qquad
%\includegraphics[width=.4\textwidth]{example-image-b}
\caption{Magnification light curves or $|F|$ as the function of time $t$ with different $w$, where the legends are similar to Figure~\ref{fig:F_w}.}
\label{fig:F_t}
\end{figure}

We now consider a specific case: a binary system composed of a supermassive lens with $M_{\rm L}=4\times10^6M_\odot$ and a rotating asymmetric NS with mass $M_{\rm S}\sim1M_\odot$ serving as a GW source. Alternatively, the GW source could be any other type of continuous emitter, such as a stellar-mass BH or a regular star, whose rotation around the supermassive BH generates continuous GWs during the early inspiral phase. For simplicity, we assume the binary system is oriented edge-on to our line of sight. The relative position of the source to the lens is given by $\eta=a\sin \omega_{\star}t$, where $a\simeq D_{\rm LS}$ represents the orbit radius, and $\omega_{\star}$ is the angular velocity of the binary. Utilizing Kepler's third law, $GM=\omega_{\star}^2a^3$, where $M=M_{\rm L}+M_{\rm S}$ is the total mass, we can determine $\omega_{\star}=\sqrt{\frac{GM}{a^3}}$ for a given semimajor axis $a$. Due to the binary's continuous rotation, the GW source is intermittently eclipsed by its companion. For our analysis, we focus on the light curve within a limited phase angle range of $-\frac{\pi}{4}<\omega_{\star}t<\frac{\pi}{4}$. Within this range, we make certain approximations. Although $D_{\rm L}$, $D_{\rm LS}$, and $D_{\rm S}$ technically vary with time $t$, for simplicity, we neglect their variations within this narrow phase range as they remain within the same order of magnitude. Clearly, $y=\frac{D_{\rm L}}{\xi_{0} D_{S}}\eta=\frac{D_{\rm L}}{\xi_{0} D_{S}}a\sin \sqrt{\frac{GM}{a^3}}t$ exhibits an approximately linear relationship with $t$. Assuming $D_{\rm L}\simeq D_{\rm S}\sim8$ kpc $\gg a=D_{\rm LS}\sim 0.1$ AU, and a lens mass of $M_{\rm L}=4\times10^6M_\odot$, we illustrate the amplification factor $F(w,y)$ as a function of time $t$. This is analogous to the light curve in electromagnetic wave microlensing. Here, we adopt a point mass model for the supermassive BH lens.

Figure~\ref{fig:F_t} depicts $|F|$ as a function of time $t$, representing the light curve. $|F|$ oscillates with time, particularly when $|t|$ is high, peaking at $t=0$ (i.e., $y=0$). The shape of this curve closely resembles $|F|$ as a function of $y$ (refer to Figure~\ref{fig:F_y}), and our findings align with those reported in \cite{2019ApJ...875..139L,2021SCPMA..6420462Z}. As in Figure~\ref{fig:F_y}, the blue dot-dashed line represents $|F_{\rm app}|$ calculated using the approximate diffraction integral, while the red solid line denotes $|F_{\rm exa}|$ computed via the exact diffraction integral. Notably, the red line consistently lies above the blue line, indicating a significant difference between the two formulas. We can draw parallels to the conclusions presented in Figure~\ref{fig:F_y}.

\subsection{Ratio $r_F$ as the function of $D_{\rm LS}$}
\begin{figure}[htbp]
\centering
\includegraphics[width=.7\textwidth]{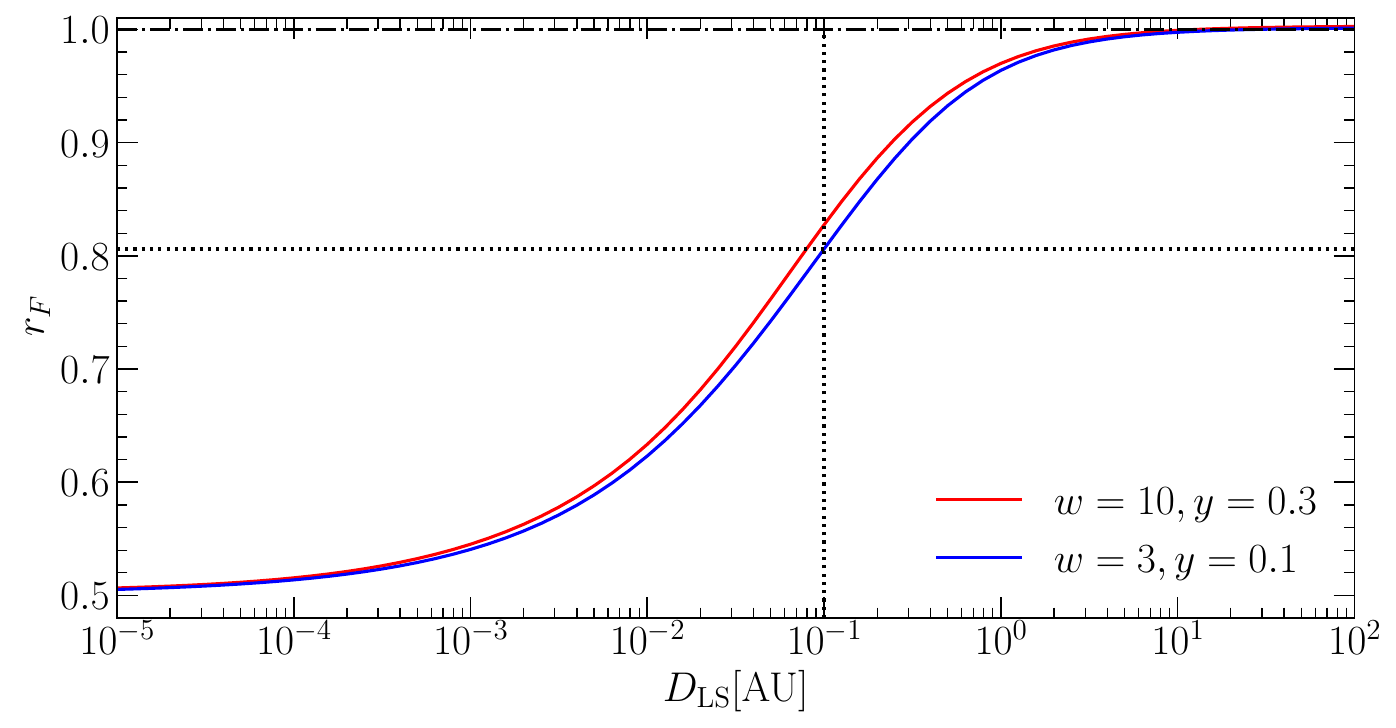}%GLnew/Plot.ipynb  &   rF_DLS.nb
\caption{Ratio $r_F$ as function of $D_{\rm LS}$. Different colors represent that they have different parameters $w$ and $y$ as shown in the lower right legends. We have chosen two groups of $w$ and $y$ values.}
\label{fig:F_D_ls}
\end{figure}

In the preceding examples, we assumed consistent lens and source distance parameters with $D_{\rm L}\simeq D_{\rm S}\sim8$\,kpc$\gg D_{\rm LS}\sim 0.1$\,AU. However, it's pertinent to inquire about scenarios where these distances are not static but variable. In this subsection, we explore the impact of lens and source distances on the amplification factor $|F|$ (as defined in Eq.~\ref{eq:F_wycos}).

Examining the expressions for $\cos\theta$ and $\cos\theta'$ (derived from Eq.~\ref{eq:F_wycos} or Eq.~\ref{eq:F_wycospolar}), it becomes evident that only three ratios — $\frac{D_{\rm L}}{\xi_0}$, $\frac{D_{\rm LS}}{\xi_0}$, and $\frac{D_{\rm S}}{D_{\rm L}}$ — influence the values of $\cos\theta$ and $\cos\theta'$. Notably, $\frac{D_{\rm L}}{\xi_0}$ is always a significant number, even when considering objects as small as the Sun (as discussed in Sec.~\ref{sec:sun}). Therefore, variations in this ratio do not significantly alter the values of $\cos\theta$ and $\cos\theta'$. In the context of binary systems, the relationship $D_{\rm L}\simeq D_{\rm S}\gg D_{\rm LS}$ remains valid. This implies that the ratio $\frac{D_{\rm S}}{D_{\rm L}}$ is approximately equal to 1, regardless of the specific distances $D_{\rm L}$ and $D_{\rm S}$. Consequently, the primary factor influencing the values of $\cos\theta$ and $\cos\theta'$ is the variation in $D_{\rm LS}$. Therefore, in this analysis, we focus primarily on the impact of changes in $D_{\rm LS}$ on the amplification factor $|F_{\rm exa}|$.

The distinction between the exact and approximate diffraction integral formulas lies in the factor $\frac{(\cos\theta+\cos\theta')\cos\theta}{2}$ within the integrand. In previous sections, we observed that the ratio of amplification factors, denoted as $\frac{|F_{\rm exa}|}{|F_{\rm app}|}$, remains largely constant at a value of $r_F$, except in the vicinities of nodes. Therefore, the ratio $r_{F}$ serves as a proxy for the divergence between the exact and approximate diffraction integral formulas.

Adopting the same parameters as listed in the (4)th-row of Table~\ref{tab:D_ratio}, with the exception of $D_{\rm LS}$, we evaluated the ratio of amplification factors $r_{F}$ as a function of the distance $D_{\rm LS}$ for various combinations of $w$ and $y$. The results are presented in Figure~\ref{fig:F_D_ls}. It's important to note that the chosen values of $w$ and $y$ are not in close proximity to any nodes, ensuring that these $r_{F}$ values are representative of more general scenarios with differing parameters.
In Figure~\ref{fig:F_D_ls}, we consider two sets of $w$ and $y$ values, which yield consistent results. For instance, when $D_{\rm LS}=0.1$\,AU, we find $r_F\simeq0.806$, aligning with our previous observations. Notably, $r_{F}$ increases from $\frac{1}{2}$ to $1$ as $D_{\rm LS}$ increases. Specifically, when $D_{\rm LS}<10^{-5}$\,AU, $r_{F}$ approaches 0.5, whereas when $D_{\rm LS}>10$\,AU, $r_{F}$ is approximately 1. This trend can be attributed to the fact that as $D_{\rm LS}$ increases, the average value of $\cos\theta'$ rises from 0 to 1, causing the factor $\frac{(\cos\theta+\cos\theta')\cos\theta}{2}$ to increase accordingly, ultimately approaching 1 when $\cos\theta'\simeq1$. In summary, significant differences between the exact and approximate diffraction integral formulas may arise when $D_{\rm LS}<10$\,AU in this particular context.

\subsection{Ratio $r_F$ as the function of $R_{\rm S}$}

\begin{figure}[htbp]
\centering
\includegraphics[width=.5\textwidth]{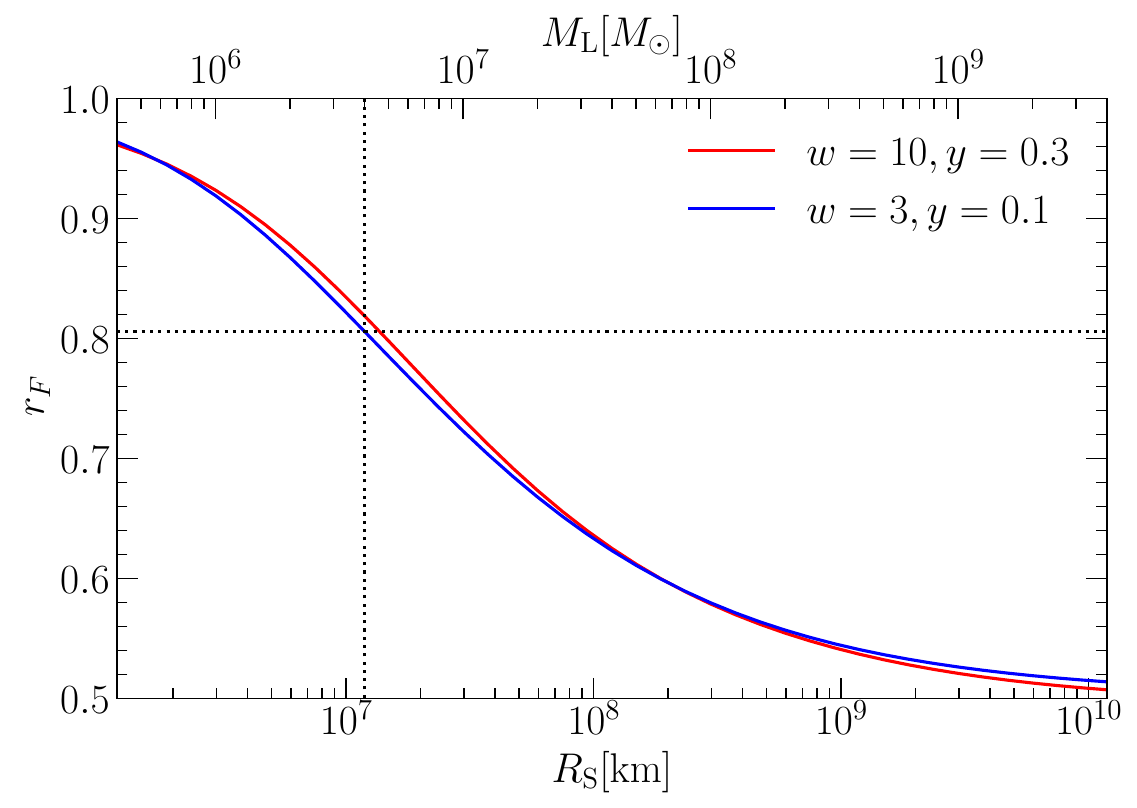}%GLnew/Plot.ipynb  &   rF_RS.nb
\caption{Ratio $r_F$ as function of $R_{\rm S}$. And the upper horizontal axis represents the corresponding lens mass $M_{\rm L}$. The legends are similar to Figure~\ref{fig:F_D_ls}. }
\label{fig:F_R_s}
\end{figure}

In addition to $D_{\rm LS}$, the Schwarzschild radius $R_{\rm S}$ of the lens also impacts $r_F$. Analogous to Figure~\ref{fig:F_D_ls}, Figure~\ref{fig:F_R_s} illustrates the ratio $r_{F}$ as a function of the Schwarzschild radius $R_{\rm S}$ (or equivalently, the lens mass $M_{\rm L}$) for various values of $w$ and $y$. In this figure, only $R_{\rm S}$ varies, while $w$ and $y$ are indicated in the legends, and all other parameters remain consistent with the (4)th-row of Table~\ref{tab:D_ratio}. If the lens has a larger $R_{\rm S}$ or, equivalently, if the lens is more massive, $r_F$ becomes smaller. Consequently, the disparity between the exact and approximate formulas becomes more pronounced. The value of $r_F$ consistently falls between 0.5 and 1. Specifically, when the lens mass $M_{\rm L}$ exceeds approximately $3\times10^{9}M_\odot$ or when $R_{\rm S}$ surpasses approximately $10^{10}$\,km, $r_F$ approaches 0.5. Conversely, for lens masses below $10^{6}M_\odot$, $r_F$ tends towards 1.

%Besides $D_{\rm LS}$,  the Schwarzschild radius $R_{\rm S}$ of lens will also influence $r_F$. Similar to Figure~\ref{fig:F_D_ls}, we also show the ratio $r_{F}$ as the function of Schwarzschild radius $R_{\rm S}$ (or lens mass $M_{\rm L}$) for different $w$ and $y$ in the Figure~\ref{fig:F_R_s}. Here only $R_{\rm S}$ is variable, $w$ and $y$ are shown in the legends,  other parameters are the same as the (4)th-row in Table~\ref{tab:D_ratio}. If $R_{\rm S}$ of lens is larger or lens is more massive, $r_F$ is smaller, thus the difference between exact and approximate formula is more significant. The value of $r_F$ is always between 0.5 and 1. When lens mass $M_{\rm L}\gtrsim3\times10^{9}M_\odot$ or $R_{\rm S}\gtrsim 10^{10}$\,km, $r_F\simeq0.5$.  When $M_{\rm L}<10^{6}M_\odot$, $r_F$ is close to 1.

\section{Discussion}
\label{sec:dis}

Firstly, it is probable that GW source-lens systems exist within our local galactic neighborhood. References like \cite{2000ApJ...545..847M,2022ApJ...939...55G} indicate the potential presence of GW sources in the Galactic center. In fact, the observation of a magnetar in this region \cite{2013Natur.501..391E,2013ApJ...770L..24K} suggests the possible existence of numerous pulsars in close proximity to the Galactic center. If the GW source originates from a rotating asymmetric NS, it could potentially be detected by ground-based GW detectors, especially the next-generation ones. Furthermore, such a GW source-lens system might constitute an extreme mass ratio inspiral (EMRI) system, whose GW emissions could be picked up by space-based GW detectors or pulsar timing arrays \cite{2022ApJ...939...55G}. Beyond the Galactic center, similar systems might also exist in neighboring galaxies like M31 and M87 (refer to Table~\ref{tab:D_ratio}). These systems provide an excellent opportunity to validate the exact diffraction integral formula.

Secondly, the magnification differences for the case that we consider are significant enough to be detectable. As discussed in \cite{2020PhRvD.101b4039M}, a relative strain difference of $\frac{\delta h}{h}>\frac{1}{\varrho}$ is detectable, where $\varrho$ is the signal-to-noise ratio (SNR) of the GW signal. For a GW signal with an SNR of 8, any relative strain difference exceeding $\frac{1}{8}$ is detectable. Our calculations reveal a substantial relative magnification difference between the exact and approximate diffraction formulas, specifically $r_F\simeq 0.806$ which translates to $\Delta h/h\simeq0.194>\frac{1}{8}$ when $D_{\rm LS}=0.1$\,AU. Referring to Figure~\ref{fig:F_D_ls}, as long as $D_{\rm LS}\lesssim0.2$\,AU, the relative magnification difference is significant enough to distinguish for a GW signal with an SNR of 8 or higher. If the lens mass $M_{\rm L}$ (or $R_{\rm S}$) varies, a significant relative magnification difference can be observed for $M_{\rm L}\gtrsim2\times 10^{6}M_\odot$ and an SNR of 8 or more (see Figure~\ref{fig:F_R_s}).

Thirdly, it should be noted that nature of GWs are tensor waves, not scalar waves. When the angle between the lens and the source is significant, the polarization tensor of the GW cannot be treated as constant. And when the source is close to lens, the spacetime is curved\cite{Dolan:2017zgu}. While our exact diffraction formula might not perfectly describe tensor wave nature in such cases, it can serve as a useful approximation, bridging the gap between approximate diffraction formulas and more complex tensor diffraction models. Future work should consider the tensor diffraction formula \cite{2019PhRvD.100f4028H} and its polarizations \cite{2022JCAP...10..095L}. He et al. \cite{2020arXiv200510485H,2021MNRAS.506.5278H,2022arXiv220501682Q,2024PhRvL.132a1401Y} have simulated GW lensing systems in their studies. To obtain more accurate lensed GW waveforms in such scenarios and to assess the accuracy of our diffraction formula, further simulations and research are warranted.

Furthermore, our precise diffraction formula can be validated not only through lensed gravitational waves but also electromagnetic wave observations. Specifically, it can be tested using lensed fast radio bursts, as exemplified in \cite{2020MNRAS.497.4956J}. If a fast radio burst source, gamma-ray burst, or a star (as a light source) is eclipsed or lensed by a supermassive black hole, it could serve as a crucial tool to verify our exact diffraction formula. While geometrical optics might prevail in the description of lensed light, it's worth noting that under the high-frequency limit, there could still exist a substantial disparity between the exact and approximate diffraction formulas. Additional research in related fields is warranted. Moreover, our exact diffraction formula is applicable to the diffraction of scalar waves, such as ultralight scalar fields with long wavelengths, providing an accurate description of such lensed wave systems.

In addition, our exact diffraction formula can be tested by not only lensed GW, but also lensed EM wave observation. For example, it can be also tested by lensed fast radio burst \cite{2020MNRAS.497.4956J}. If a fast radio burst source, gamma ray burst or a star (light source) eclipsed or lensed by a supermassive black hole could become an important probe to test our exact diffraction formula. Although it could be dominant by geometrical optics for lensed light, there could not exist significant difference between exact and approximate diffraction formulas under high frequency limit \cite{Dolan:2017zgu}. More investigations need to be done in related fields. What is more, our exact diffraction formula can be also used to describe the diffraction of a scalar wave, e.g. an ultralight scalar field with long wavelength. For scalar wave, it is accurate to describe such lensed wave system with this formula.
%\cite{1995ApJ...442...67U}

\section{Conclusions}
\label{sec:con}

In previous work \cite{2020PhRvD.102l4076G}, we introduced an exact and comprehensive diffraction integral formula that encompasses large-angle diffraction scenarios. Our investigations reveal that when GWs are lensed by a supermassive BH, the disparities between the exact and approximate diffraction integrals can be substantial. Consequently, GW sources lensed by supermassive BHs offer an ideal testing ground for our exact diffraction formula. Such binary systems are likely to exist in the centers of nearby galaxies \cite{2022ApJ...939...55G}, making them promising targets for detection by next-generation GW detectors, space-based observatories, pulsar timing arrays, or detectors operating in other frequency bands.

Furthermore, we have quantified the disparities between the exact and approximate diffraction integrals across various $w$ and $y$ values. Generally, the module of amplification factor derived from the general diffraction formula is smaller than that obtained from the approximate method. Their ratio, denoted as $\frac{|F_{\rm exa}|}{|F_{\rm app}|}$, remains largely constant at a value of $r_F$, except at certain nodal points.
Considering a specific case where the lens-source distance is $D_{\rm LS}=0.1$\,AU and the lens mass is $M_{\rm L}=4\times10^{6}M_\odot$, the proportionality factor is approximately $r_F\simeq0.804$. This factor $r_F$ varies depending on $D_{\rm LS}$ and the lens mass $M_{\rm L}$ (or equivalently, the Schwarzschild radius of the lens $R_{\rm S}$). Assuming a constant lens mass of $M_{\rm L}=4\times10^{6}M_\odot$, we observe that when $D_{\rm LS}<10^{-5}$\,AU, $r_{F}\simeq0.5$. As $D_{\rm LS}$ increases from $10^{-5}$\,AU to 10\,AU, $r_{F}$ progressively rises from 0.5 to 1. Beyond $D_{\rm LS}>10$\,AU, $r_{F}$ is stabilized around 1. Significant disparities between the exact and approximate diffraction integral formulas emerge only when $D_{\rm LS}<10$\,AU. Specifically, when $D_{\rm LS}<0.2$\,AU, these differences become substantial enough to be detectable for any GW signal with a SNR ($\varrho$) exceeding 8.
Keeping $D_{\rm LS}$ fixed at 0.1\,AU, we note that $r_F$ decreases as $M_{\rm L}$ increases. Once $M_{\rm L}$ surpasses $3\times10^{9}M_\odot$, $r_F$ is stabilized at approximately 0.5. For lens masses $M_{\rm L}>2\times10^6M_\odot$, the differences between the exact and approximate formulas become significant enough to be discernible for GW signals with a $\varrho$ greater than 8.
In conclusion, our exact general diffraction formula holds promise to be verified for future GW detection efforts, particularly in scenarios involving supermassive BHs.

\appendix
\section{Phase difference in lensing}
\label{sec:phase}

\begin{figure}[htbp]
\centering
\includegraphics[width=.4\textwidth]{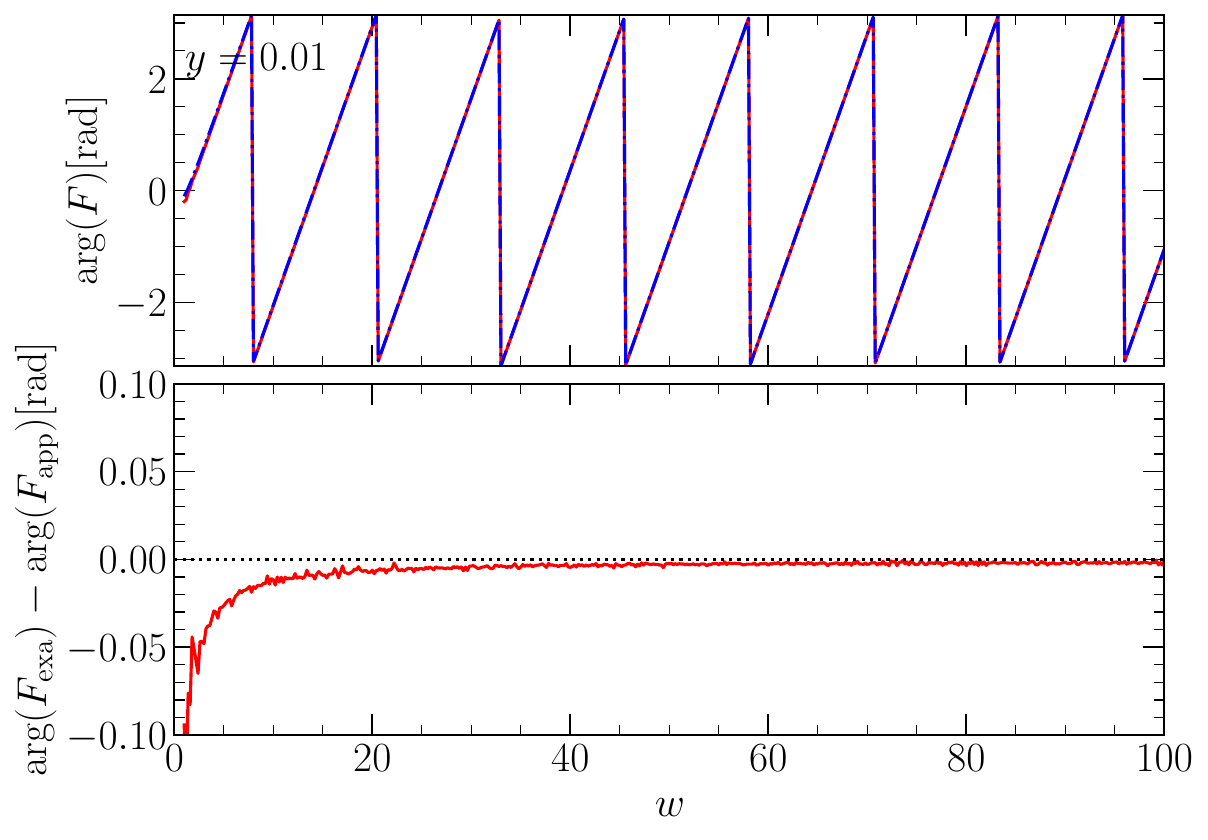}
\caption{Phase of amplification factor $\arg(F)$ from exact (red solid line) and approximate (blue dot-dash line) diffraction integrals and their phase difference (bottom panel).} 
\label{fig:phase}
\end{figure}

Analogous to Figure~\ref{fig:F_w}, Figure~\ref{fig:phase} illustrates the phase of the amplification factor, $\arg (F)$, as a function of frequency $w$. For simplicity, we assume $\phi_m(y)\equiv0$ in calculating the absolute phase $\arg (F)$. While the absolute phase is inconsequential, the phase difference between various factors holds significance. The phase difference, denoted as $\Delta \Phi=\arg(F_{\rm exa})-\arg(F_{\rm app})$, between the exact and approximate formulas is remarkably small, with an absolute value typically below 0.1\,rad. As the frequency $w$ approaches infinity, this phase difference tends to zero: $\Delta \Phi=\arg(F_{\rm exa})-\arg(F_{\rm app}) \rightarrow0$. Consequently, the phase difference remains insignificant across various $w$ and $y$ values, at least for frequencies $w<100$.

%Similar to Figure~\ref{fig:F_w}, we also show phase of amplification factor $\arg (F)$ as the function of frequency $w$ in Figure~\ref{fig:phase}. For simplicity, we assume $\phi_m(y)\equiv0$ in the calculation of value of absolute phase $\arg (F)$. The absolute phase is not important, but the phase difference between them matters. The phase difference $\Delta \Phi=\arg(F_{\rm exa})-\arg(F_{\rm app})$ between the exact formula and approximate one is very small. The absolute value of phase difference is basically less than 0.1\,rad. And when $w\rightarrow \infty$, phase difference $\Delta \Phi=\arg(F_{\rm exa})-\arg(F_{\rm app}) \rightarrow0$. Thus the phase difference is not significant for all kinds of values of $w$ and $y$ at least for $w<100$. 

\acknowledgments
Zhoujian Cao acknowledges the National Key Research and Development Program of China (Grant No. 2021YFC2203001).
Xiao Guo acknowledges the fellowship of China National Postdoctoral Program for Innovative Talents (Grant No. BX20230104).

% Bibliography

%% [A] Recommended: using JHEP.bst file
\bibliographystyle{JHEP}
\bibliography{biblio.bib}

%% or
%% [B] Manual formatting (see below)
%% (i) We suggest to always provide author, title and journal data or doi:
%% in short all the informations that clearly identify a document.
%% (ii) please avoid comments such as "For a review'', "For some examples",
%% "and references therein" or move them in the text. In general, please leave only references in the bibliography and move all
%% accessory text in footnotes.
%% (iii) Also, please have only one work for each \bibitem.

\end{document}